\documentclass[a4paper,12pt]{article}

\usepackage{amssymb,amsmath,mathbbol,mathrsfs}
\usepackage[usenames,dvipsnames]{color}
\usepackage[version=3]{mhchem}
\usepackage{hyperref}
\usepackage[table]{xcolor}
\usepackage{multirow,bigdelim}
\usepackage{stmaryrd}
\usepackage{authblk}
\usepackage{framed}
\usepackage{empheq}
\usepackage{slashed}
\usepackage{hyphenat}
\usepackage{cite}

\usepackage{tikz}
\usepackage{tikz-3dplot}
\usetikzlibrary{calc}
\usetikzlibrary{shapes.geometric,patterns} 
\usetikzlibrary{tikzmark}

\usepackage{marginnote}    

\usepackage[left=0.5in,right=1in,top=1in,bottom=1in,marginparwidth=1.5in, marginparwidth=2.5cm, marginparsep=.2cm]{geometry}       

\usepackage{sectsty}
\allsectionsfont{\sffamily\mdseries\upshape} 
\usepackage{tocloft}

\makeatletter
\renewenvironment{abstract}{%
    \if@twocolumn
      \section*{\abstractname}%
    \else 
      \begin{center}%
        {\bfseries\sffamily\abstractname\vspace{\z@}}
      \end{center}%
      \quotation
    \fi}
    {\if@twocolumn\else\endquotation\fi}
\makeatother

5

\hypersetup{
	colorlinks=true,         
	linkcolor=MidnightBlue,          
	citecolor=BrickRed,        
	urlcolor=MidnightBlue            
}

\newcommand{\be}{\begin{equation}}
\newcommand{\ee}{\end{equation}}
\newcommand{\bes}{\begin{eqnarray}}
\newcommand{\ees}{\end{eqnarray}}

\renewcommand{\tilde}{\widetilde}

\renewcommand{\d}{{\mathrm{d}}}

\newcommand{\DD}{{\mathscr{D}}}
\renewcommand{\b}{{\mathrm{b}}}

\newcommand{\Ad}{{\mathrm{Ad}}}
\newcommand{\pp}{{\partial}}

\newcommand{\su}{{\mathfrak{su}}}
\renewcommand{\bar}{\overline}

\newcommand{\ra}{\rangle}

\newcommand{\SU}{\mathrm{SU}}
\newcommand{\SO}{\mathrm{SO}}
\newcommand{\so}{\mathfrak{so}}

\newcommand{\act}{\triangleright}
\renewcommand{\b}{\overline}

\newcommand{\cint}{{\int\kern-.87em{<}}}
\newcommand{\sint}{{\int\kern-.75em{\sim}}}
\newcommand{\fint}{{\int\kern-1.00em{\int}}}

\newcommand{\bb}{\mathbb}

\newcommand{\tr}{\mathrm{Tr}}

\newcommand{\q}{\quad}

\def\cA{{\mathcal{A}}}
\def\cB{{\mathcal{B}}}
\def\cF{{\mathcal{F}}}
\def\cG{{\mathcal{G}}}
\def\rd{{\mathrm{d}}}

\def\q{\quad}

\def\nn{\nonumber}

\let\oldmarginpar\marginpar
\renewcommand\marginpar[1]{\oldmarginpar{\color{red}\raggedright\footnotesize #1}}

\title{\sffamily 
 Quantum geometry from higher gauge theory
 }
\author[1,2]{\sffamily Seth K. Asante\thanks{sasante@perimeterinstitute.ca}}
\author[1]{\sffamily Bianca Dittrich\thanks{bdittrich@perimeterinstitute.ca}}
\author[2]{\sffamily Florian Girelli\thanks{florian.girelli@uwaterloo.ca}}
\author[1]{\sffamily Aldo Riello\thanks{ariello@perimeterinstitute.ca}}
\author[2]{\sffamily Panagiotis Tsimiklis\thanks{ptsimiklis@uwaterloo.ca}}
\affil[1]{\small Perimeter Institute for Theoretical Physics, 31 Caroline St. N., Waterloo, ON N2L2Y5, Canada}
\affil[2]{\small Department of Applied Mathematics, University of Waterloo, 200 University Avenue West, Waterloo, Ontario, Canada, N2L 3G1}

\begin{document}
\maketitle

\abstract{ \small

Higher gauge theories play a prominent role in the construction of 4d topological invariants and have been long ago proposed as a tool for 4d quantum gravity. The Yetter lattice model and its continuum counterpart, the BFCG theory, generalize BF theory to 2-gauge groups and -- when specialized to 4d and the Poincar\'e 2-group -- they provide an exactly solvable topologically-flat version of 4d general relativity. The 2-Poincar\'e Yetter model was conjectured to be equivalent to a state sum model of quantum flat spacetime developed by Baratin and Freidel after work by Korepanov (KBF model). This conjecture was motivated by the origin of the KBF model in the theory of 2-representations of the Poincar\'e 2-group. Its proof, however, has remained elusive due to the lack of a generalized Peter-Weyl theorem for 2-groups.

In this work we prove this conjecture. Our proof avoids the Peter-Weyl theorem and rather leverages the geometrical content of the Yetter model. Key for the proof is the introduction of a kinematical boundary Hilbert space on which 1- and 2-Lorentz invariance is imposed. Geometrically this allows the identification of (quantum) tetrad variables and of the associated (quantum) Levi-Civita connection. States in this Hilbert space are labelled by quantum numbers that match the 2-group representation labels. 

Our results open exciting opportunities for the construction of new representations of quantum geometries. Compared to loop quantum gravity, the higher gauge theory framework provides a quantum representation of the ADM-Regge initial data, including an identification of the intrinsic and extrinsic curvature. Furthermore, it leads to a version of the diffeomorphism and Hamiltonian constraints that acts on the vertices of the discretization, thus providing a prospect for a quantum realization of the hypersurface deformation algebra in 4d.

 }


\newpage
{\hypersetup{	linkcolor=black, }\tableofcontents}
\begin{center}
	\rule{8cm}{0.4pt}
\end{center}
{\hypersetup{	linkcolor=MidnightBlue }}

\newpage


\newpage
\section{Introduction and summary of the results\label{sec:intro}}

The reformulation of general relativity in terms of gauge variables, in particular the introduction of the Ashtekar variables for canonical gravity \cite{Ashtekar:1986yd}, has led to the development of loop quantum gravity (LQG) -- for a recent review, see  \cite{Ashtekar:2017yom}.
 Among the many accomplishments of LQG, one counts different realizations of quantum geometry \cite{Ashtekar:1991kc, Ashtekar:1993wf, Koslowski:2011vn, Dittrich:2014wpa, Bahr:2015bra, Delcamp:2016yix, Drobinski:2017kfm,Dittrich:2016typ,Dittrich:2017nmq,Dupuis:2013haa}. 
A quantum geometry realization is constituted  by a Hilbert space encoding boundary metric degrees of freedom and their conjugated momenta, as well as by a family of quantum geometric operators acting on them. 
 Different realizations can be thought of as inequivalent phases attained by the geometric degrees of freedom.

The  Palatini-Plebanski formulation of four-dimensional general relativity \cite{Plebanski:1977zz} provides a covariant gauge theoretic description closely related to canonical LQG.
In three dimensions the related   Einstein-Cartan formulation puts general relativity in the form of a BF topological field theory \cite{Horowitz:1989ng}.
In four dimensions, on the other hand, the  Palatini-Plebanski formulation rephrases general relativity as emerging from a BF theory by imposition of the so-called simplicity constraints. 
This perspective has led to the development of covariant state-sum models for four-dimensional quantum gravity known as ``spinfoam'' models \cite{Perez:2012wv, Freidel:2007py, Engle:2007wy, Barrett:1997gw,Reisenberger:1996pu}, as well as to the introduction of canonical realization of quantum geometry based on BF theory \cite{Dittrich:2014wpa, Dittrich:2014wda, Bahr:2015bra}.

In three spacetime dimensions, this gauge-theoretic formulation based on the Einstein-Cartan formulation has been particularly successful \cite{Rovelli:1993kc, Noui:2004iy, Freidel:2004vi, Barrett:2008wh, Dittrich:2018xuk}.
The theory is readily rephrased as a topological BF theory, and its quantum partition function takes the form of a state sum model written in terms of representation theoretic objects for the corresponding gauge group assigned to a simplicial decomposition of the three-manifold.
This state sum, known as the Ponzano-Regge model, was first proposed back in 1968 \cite{PonzanoRegge}.  
In a more modern language, the Ponzano-Regge model and its closely related generalization with cosmological constant \cite{Turaev:1992hq,Mizoguchi:1991hk} can be understood in terms in terms of category theory, and in particular the category of group representations \cite{Turaev:1992hq, Barrett:1993ab}. 
The identities ensuring the topological invariance of the model can then be reinterpreted as the coherence relations for the category of group representations.
In this sense category theory is well adapted to describe three-dimensional topological models \cite{Baez:2005qu,Baez:2009as,Baez:2010ya}.

In four dimensions, a similar relation holds between four-dimensional topological invariant models and 2-categories \cite{Baez:1995xq,Baez:2005qu,Baez:2010ya,Delcamp:2018wlb}. 
Concrete topologically invariant state sum models can then be constructed from the 2-category of 2-representations of 2-groups, e.g.  \cite{Baratin:2014era}.
For these reasons, 2-categories and ``higher'' gauge theories\footnote{The name makes reference to a ``higher categorification'' of the structures proper of gauge theories, hence generalizing the latter to 2- or higher gauge theories.} have been since long proposed to play an important role for four-dimensional quantum gravity \cite{Crane:1995qj,Smolin:1995vq,Girelli:2007tt,Mikovic:2011si,Mikovic:2018vku}.
In direct generalization of BF theory, so-called BFCG theory \cite{Girelli:2007tt} is a topological field theory displaying not only a 2-gauge symmetry structure but also a four-dimensional geometric interpretation: in addition to a 1-connection and a bivector field (also present in 4d BF theory), BFCG features a 2-connection and a tetrad field (encoded in the ``CG'' part of 4d BFCG theory). 
 This theory -- analyzed from a ``physicist's perspective\footnote{Technical features of higher gauge theories will be introduced if and when they are needed. Any direct discussion of the ``abstract nonsense'' will be avoided and the reader referred to the broad existing literature.} -- will be the focus of this article, and the presence of the tetrad field will play a central role.

Of course, four-dimensional gravity is not topological: constraints have to be imposed in order to obtain gravity from BFCG theory. 
However, this task has been severely hindered by the lack of a (Peter-Weyl) transformation between functions on 2-groups and functions of the corresponding 2-representation theoretic data. 

Nevertheless, a 2-representation theoretic partition function tied to 4d simplicial geometry has been constructed by Baratin and Freidel \cite{Baratin:2006gy, Baratin:2009za, Baratin:2014era}.
We name the corresponding topological state sum model the KBF model,\footnote{In a previous publication by two of us (SA and BD) \cite{Asante:2018kfo}, the KBF model was referred to as the BFR model, the R standing for Regge (as we will see, the model somewhat resembles Regge calculus).} since its topological invariance relies on fascinating identities first studied by Korepanov \cite{Korepanov:2002tp,Korepanov:2002tq,Korepanov:2002tr}.
 This model can be interpreted as describing a (non-physical) theory of quantum flat spacetime.
Although conjectured, the relationship between the KBF model and BFCG theory -- or its lattice analogue, the Yetter model\footnote{The Yetter model is usually formulated for finite 2-groups. This avoids the emergence of divergencies. In this paper, working on a single 4-simplex, we will generalize the Yetter model to the (Euclidean) Poincar\'e 2-group. 
} \cite{yetter1993tqft} -- has been elusive so far.

In this work, we present a way to derive the KBF model from the BFCG-Yetter model. 
This derivation will {\it not} employ a  Peter-Weyl like transformation, at least not on the full (2-group) configuration space. 
Instead, we will first impose geometrically motivated constraints, and then apply a transformation to the reduced configuration space. 
The resulting variables remarkably match the relevant 2-representation theoretic data, and the resulting amplitude that of the KBF model. 

 Beside possibly providing cues on how to develop a simplicial ``2-Peter-Weyl'' transformation, the main asset of our derivation is to highlight the quantum geometric content of the BFCG-Yetter theory through its relation to the KBF model that -- as we mentioned above -- readily admits the interpretation of a theory for a flat 4d quantum spacetime. 
In this sense, the present work provides a precious starting point to build new realization of quantum geometry that are genuinely 4d.

At this purpose, a crucial advantage 4d BFCG theory has over 4d BF theory, is that among its fundamental variables it includes the tetrad field -- which in turn allows to define a Levi-Civita connection.  
This has to be contrasted to 4d spinfoam models based on 4d BF theory: here, the tetrad field has to arise from the BF bivector $B$, through the imposition of so-called simplicity constraints -- a process that has proven to be fraught.

The other new variable present in BFCG theory with respect to BF theory is a 2-connection.
As much as the bivector field is the momentum conjugated to the 1-connection (in both theories), the 2-connection provides the momentum conjugate to the tetrad field.
In BFCG the 2-connection and the bivector field do not have a specific geometric interpretation and rather play the role of Lagrange multipliers for the torsionless condition of the 1-connection relative to the tetrad field, and for the 1-flatness condition for the 1-connection, respectively.
In particular, the 2-flatness constraint for the 2-connection takes in the discrete a very simple form (a sum over momenta\footnote{This is suggestively remindful of \cite{Freidel:2019ees}.}) and, through its action on the tetrad field, it can be readily interpreted as a combination of the diffeomorphism and Hamiltonian constraint for its action on the tetrad field.\footnote{What the 2-flatness constraint does {\it not} do is to generate diffeomorphism transformations for the 1-connection. More on this in the conclusion, section \ref{outlook}.}
More precisely, the 2-flatness constraint generates spacetime translations for the quantum-flat simplicial geometry -- thus generalizing for the first time to four dimensions what is understood as the action of the diffeomorphism and Hamiltonian constraints in three dimensions \cite{Freidel:2002dw,Baratin:2011tg,Bonzom:2013tna}.

Therefore, the 4d BFCG model based on the Poincar\'e 2-group, through its precise relation to the KBF model described in this work, has the potential to lead to a new tetrad-based quantum geometry representation in $(3+1)$-dimensions.  
This will require a further imposition of constraints, which relate the 2-connection and the tetrads to the 1-connection and the bivector fields, respectively.
 This problem is conceptually analogous to the problem of imposing the simplicity constraints in BF theory, but technically quite different. 
Thus we hope that the new tools presented in this work will be able to bring new perspectives on the construction of genuinely four-dimensional spinfoam models.

We conclude the introduction by sketching the three main innovations that allowed us to make significant progress, with respect to the previous literature, towards proving the correspondence between the KBF and the BFCG-Yetter model.
\begin{enumerate}
\item 
The first  innovation is the introduction of {\it boundaries}: in other words, we focus on the BFCG-Yetter amplitude of a single four-simplex, bounded by 5 tetrahedra. 
This immediately leads us to consider boundary states for this amplitude that impose specific quantum boundary conditions.\footnote{During the final stages of the preparation of this work, we became aware that A. Baratin had hinted at the importance of considering a single building block and its boundary. He also emphasized that the best 3d analogue to the BFCG-Yetter-KBF model is 3d BF theory for the 3d Poincar\'e-Euclidean group discretized as a 1-gauge theory. This material is unpublished, but was presented at a conference \cite{AristideTalk}. }
\item 
Inspired by spin-network states relevant for relating the Ponzano-Regge amplitude of a tetrahedron to 3d BF theory, we introduce (Fourier-transformed) states that solve the 1- and 2-Gauss constraints of the BFCG theory and  leave 1- and 2-flatness unconstrained. 
We call these states {\it G-networks} (G stands for Gauss). 
This is our second  innovation.
\item 
In the construction of G-networks, it is crucial to recognize that the 2-Gauss constraint, which has the interpretation of torsion freeness for the 1-connection, in the discrete context is made of two pieces: one that imposes closure of triangles, and the other that imposes ``{\it edge simplicity}'' \cite{Dittrich:2008ar}.
Whereas the first piece emerges naturally in the construction of G-networks, the second piece at first sight appears ``hidden'' in the 1-flatness (or fake-flatness in the nomenclature of \cite{Breen:2001ie,Baez:2004in}) constraint. 
Recognizing this fact is our third innovation.
\end{enumerate}

~\\
This article is organized as follows.
We start by concisely reviewing a few relevant topics. In section \ref{sec:BF2PR}, we review the relation of 3d BF theory and the Ponzano-Regge model. We will put emphasis on those ideas and techniques that will turn useful in the 4d case. We then briefly review  in section \ref{sec:BFCGYKBF} BFCG theory and its discretization into the Yetter model, as well as the KBF model.
This is followed, in section \ref{sec:4}, by a discussion of the appearance of tetrad variables in the BFCG-Yetter model.
In this context, we will focus in particular on the derivation of the torsion-freeness and edge-simplicity constraints, and on how their imposition restricts the connection to be Levi-Civita.  
The main section of this work is section \ref{sec:5}, where we detail the structures of the (discrete) BFCG-Yetter phase-space associated to the boundary of a 4-simplex, as well as its quantization. Specifically, we provide a discretization of all the relevant constraints and discuss their classical and quantum actions.  This will allow us to introduce G-networks in section \ref{sec:gnetwork} as the solutions to a certain subset of the constraints. 
We will hence show that a basis of G-networks is parametrized by the same labels as the 2-Poincar\'e group irreducible representations, a fact that will finally allow us to derive, in section \ref{KBFfromBFCG}, the KBF model from the BFCG-Yetter model. 
We conclude in section \ref{outlook} with a discussion of the results obtained as well as an outlook on future work.

\section{Three dimensions: from BF theory to the Ponzano-Regge state sum\label{sec:BF2PR}}
 
In this section, we sketch how our strategy can be applied to $\SU(2)$ BF theory in three dimensions to obtain the Ponzano-Regge state sum.
The goal of this section is solely to exemplify our strategy in a simpler context, before applying it to the derivation of the KBF model from BFCG-Yetter model. 

The $\SU(2)$ BF amplitude of a compact and closed 3-manifold $M_3$ is
\be
Z_{BF}(M_3) = \int \DD B \DD A \, e^{i \int_{M_3} \tr( B\wedge F[A] )},
\label{eq:Z_BF}
\ee
where $B$ is a $\su(2)$-valued 1-form and $A$ an $\su(2)$-valued connection 1-form. 
The BF action is invariant under ``{\it shift}'' and ``{\it Lorentz}'' gauge transformations:
\be
\text{ Lorentz: }
\begin{cases}
A \mapsto A + \d_A \xi\\
B \mapsto B + [B,\xi]
\end{cases}
\quad\text{and}\quad
\text{shift: } 
\begin{cases}
A \mapsto A\\
B \mapsto B + \d_A \lambda
\end{cases}
\ee 
where the (infinitesimal) gauge parameters $\xi$ and $\lambda$ are $\su(2)$-valued 0-forms.

The equations of motion of BF theory encode the flatness of $A$, $F[A] \equiv \d A + A \wedge A =0$, and the (covariant) constancy of $B$, $\d_A B =0$.
In topologically trivial domains, they imply that on-shell $A$ and $B$ are pure gauge. This is a sign of the topological nature of the theory. 

Diffeomorphism symmetry arises, on-shell of the equations of motion, as a combined action of Lorentz and shift symmetry with field-dependent parameters  \cite{Horowitz:1989ng}. Explicitly: given an infinitesimal diffeomorpshism $X\in\mathfrak X^1(M_3)$, we define the Lorentz and shift gauge parameters
\be
\Big( \xi_X = \iota_X A,\quad \lambda_X = \iota_X B \Big)  
\ee
and find
\be
\text{diffeos: }
\begin{cases}
A \mapsto A + \pounds_X A  = A + \iota_X F + \d_A \xi_X\\
B \mapsto B + \pounds_X B = B + \iota_X (\d_A B) + [B, \xi_X]  + \d_A \lambda_X
\end{cases}.
\label{eq:3dBFdiffeo}
\ee
Notice that, whereas $\xi_X$ only ``rotates" $B$, $\lambda_X$ changes its norm.
In three dimensional quantum gravity this fact has been used to reconstruct the action of diffeomorphisms at the level of the Ponzano-Regge state sum \cite{Freidel:2002dw}. 

Neglecting Lagrange multipliers, the phase space of BF theory on a Cauchy hypersurface is given by the pullbacks of $A$ and $B$, $a=\iota^*A$ and $b=\iota^*B$. These are Poisson-conjugated kinematical variables.
Pulled-back on the Cauchy hypersurface, the equations of motion become the {\it flatness} and {\it Gauss} constraints, that Poisson-generate the phase space analogue of shift and Lorentz symmetry, respectively: 
\begin{subequations}
\begin{alignat}{3}
\text{flatness: } & F[a] \approx 0 &&\q\leadsto\q \text{shift} \\
\text{Gauss: } & \d_a b \approx 0 &&\q\leadsto\q \text{Lorentz}.
\end{alignat}
\end{subequations}

Identifying $\su(2)$ with $\bb R^3 \cong \mathrm T_pM_3$, and thus $B^a_\mu$ with the triad $e^a_\mu$, and $A^a_\mu$ with the spin-connection $\omega^{ab}_\mu = \epsilon^{ab}{}_c A^c_\mu$, allows us to interpret $Z_{BF}$ as the partition function of three-dimensional quantum gravity in the first order  formalism. 
In this theory, the equation of motion $\d_A B = 0$ is reinterpreted as a torsionless condition generating rotations in the tangent planes of $M_3$ (hence the name Lorentz transformations); flatness becomes Riemann flatness, which in 3 dimensions is equivalent to Ricci flatness, that is to the Einstein field equations in vacuum. The shift symmetry is in a certain sense equivalent to diffeomorphism symmetry   \cite{Freidel:2002dw, Baratin:2011tg}.

Integrating out the $B$ field from \eqref{eq:Z_BF} and then discretizing the theory on a simplicial discretization $\Delta_3$ of $M_3$, one obtains a lattice version of BF theory that (formally) reads\footnote{First discretizing and then integrating out also leads to the same result.}
\be
Z_{BF} (\Delta_3) = \int \DD H_{l} \, \prod_{f \in \Delta_3^*} \delta\left( {\overleftarrow\prod}_{l: l\in \pp f}H_{l}^{\epsilon(l|f)} \right)
\label{eq:Z_BFdiscrete}
\ee
where: 
\begin{itemize}
\item $\Delta_3^*$ is the Poincar\'e dual of $\Delta_3$, $l$ and $f$ label the (oriented) links and faces of $\Delta_3^*$ respectively (see table \ref{tab:Delta3});
\item $H_{l}\in \SU(2)$ are parallel transports (holonomies) obtained by smearing $A$ along $l$, $H_{l} = \mathrm P\exp \int_{l} A$;
\item $\DD H_{l} := \prod_{l\in\Delta^*_3} \d H_{l}$ is a product of Haar measures over the $H_{l}$; 
\item $\epsilon(l|f)=\pm 1$ captures the relative orientation of $l$ and $f$, so that the expression in parenthesis is a discrete version of the curvature: $ \overleftarrow{\prod}_{l: l\in \pp f }H_{l}^{\epsilon(l|f)}  \sim \bb 1 + \int_f F[A] + \dots$.
\end{itemize}

\begin{table}[t]
\begin{center}
\begin{tabular}{|c|c|c|c|c|}
\hline
cell dim     & 0       &1      &2        & 3               \\ \hline
$\Delta_3$                   & vertex $v$            & edge $e$              & triangle $t$          & tetrahedron $\tau$    \\ \hline
\hline
$\Delta^*_3$                 & bubble $b$              & face $f$          & link $l$             & node $n$              \\ \hline
cell codim                   & 3                     & 2                     & 1                     & 0                           \\ \hline
\end{tabular}
\caption{Notation employed to denote cells of the three-dimensional simplicial complex $\Delta_3$ and its Poincar\'e dual $\Delta_3^*$.}\label{tab:Delta3}
\end{center}
\end{table}

The Ponzano-Regge model can be obtained from \eqref{eq:Z_BFdiscrete}, by dualization \cite{Barrett:2008wh}. One starts by expanding the $\delta$'s over the $\SU(2)$ characters in the $j$-th representation $V^j$ according to the Peter-Weyl formula
\be
\delta(G) = \sum_j d_j \chi^j(G)
\qquad \text{where} \qquad
d_j = \mathrm{dim} V^j = 2j+1.
\label{eq:PW}
\ee 
This associates a spin $j_{f}$ to each $f\in\Delta_3^*$ or, equivalently, a spin $j_e$ to each edge $e\in\Delta_3$. 
This spin represents the (discrete quantum) length of the edge $e\in\Delta_3$.
Second, one integrates out the $H_{l}$'s by means of identities in recoupling theory. Since $\Delta_3$ is a triangulation of $M_3$, each $H_{l}$ appears precisely three times in \eqref{eq:Z_BFdiscrete}. Thus, by means of recoupling theory one obtains products of $(3j)$ symbols associated to the triangles of $\Delta_3$.
Finally, one observes that the so-obtained expression can be re-arranged to give the Ponzano-Regge state sum
\be
Z_{BF}(\Delta_3) \equiv Z_{PR} (\Delta_3)= \sum_{\{ j_e\} } \prod_{e\in\Delta_3} (-1)^{2{j_e}} d_{j_e} \prod_{t \in \Delta_3} (-1)^{\sum_{e:e \in t} j_e} \prod_{\tau\in\Delta_3} \{ 6 j_e \}_\tau
\label{eq:Z_PR}
\ee
where $t$ and $\tau \in \Delta_3$ are respectively the triangles and tetrahedra of $\Delta_3$, and $\{ 6j_e\}_\tau$ is a specific contraction of four $(3j)$ symbols reflecting the combinatorics of a tetrahedron $\tau$. 
In particular the 6 spins $j_e$ entering this expression correspond to the 6 edge lengths of $\tau$. 
This geometric interpretation is confirmed by the asymptotic formula\footnote{This equation holds if the 6 edge lengths $j_e$ define a geometric tetrahedron, i.e. if their Caley-Menger determinant is positive (triangular inequalities must be satisfied for the $\{6 j_e\}$ not to vanish), otherwise the amplitude is exponentially suppressed by a factor corresponding to a Lorentz-signature Regge action \cite{BarrettFoxon} -- but cf. the end of section 5.1 of \cite{AquilantiHaggardLittlejohn} for a critique of the Lorentzian interpretation.} 
\cite{PonzanoRegge, Roberts1999, AquilantiHaggardLittlejohn}
\be
\{ 6j_e \}_\tau \stackrel{j_e\gg1}{\sim} \frac{1}{\sqrt{12\pi |\mathrm{Vol}_\tau}|} \cos\Big(  S^3_R + \tfrac{\pi}{4} \Big)
\quad\text{where}\quad
S^3_R = \sum_{e\in\tau} j_e \Theta_e (\{j_e\}) ,
\label{eq:Regge_asympt}
\ee
where $|\mathrm{Vol}_\tau|$ is the volume of the tetrahedron of edges $\{j_e\}$, and $S^3_R$ is its 3-dimensional Regge action. 
This is a discretization of the 3-dimensional Einstein-Hilbert action with Gibbons-Hawking-York boundary conditions on a tetrahedron \cite{Regge:1961,Hartle:1981cf}. It consists of the lengths of the triangulation edges $j_e$ multiplied by the corresponding exterior dihedral angles $\Theta_e$, calculated as functions of the edge lengths themselves.\footnote{A Lorentzian version also exists \cite{Freidel:2000uq,Girelli:2015ija}.} 

In general dimensions, the Regge action $S_R^d$ for a $d$-dimensional simplex, is given by the product of the volume of the \textit{co}dimension 2 hinges\footnote{These are the cells of $\Delta_d$ dual to the faces $f\in\Delta_d^*$. Notice that the discrete analogue of the curvature, as measured by a holonomy, resides at $f$ independently of the dimension $d$.} times the associated (hyper-)dihedral angles -- {\it with both factors considered as functions of the edge-lengths of the triangulation $\Delta_d$.} In four dimensions:
\be
S_R^4 = \sum_{t} A_t(\{\ell_e\}) \Theta_t(\{\ell_e\}).
\label{eq:Regge_d}
\ee

In the higher-gauge theoretical case of the BFCG model, the higher categorical analogues of the Peter-Weyl and Plancharel theorems are not known. Therefore, one cannot rely on equation \eqref{eq:PW} nor on recoupling theory to go from the analogue of \eqref{eq:Z_BFdiscrete} to that of \eqref{eq:Z_PR} -- which, as we will see, are the BFCG-Yetter and KBF models respectively.

For this reason, we want to explore a different strategy\footnote{Another example of this strategy of using boundary state to better understand  the theory  is \cite{Haggard:2014xoa, Haggard:2015kew, Haggard:2015yda} which aimed at building a four dimensional spinfoam model with a cosmological constant.} to obtaining \eqref{eq:Z_PR} from \eqref{eq:Z_BFdiscrete}.
Hence, we focus on a single tetrahedron $\tau$, and consider the corresponding triangulation with boundary  $\pp \Delta_ 3 =\pp \tau$. 
The boundary $\pp \tau$ consists of four triangles, and its Poincar\'e dual is (again) a tetrahedral graph with four trivalent faces. Call it $\Gamma = (\pp \tau)^*$.
To $\Gamma$ we associate a boundary state $\Psi_\Gamma$  \cite{OLoughlin:2000yww, Bonzom:2015ans, Dittrich:2017hnl, Dittrich:2017rvb, Dittrich:2018xuk, Riello:2018anu}.   
In the Schr\"odinger representation, the amplitude of $\Psi_\Gamma$ is
\be
Z_{BF}(\tau|\Psi_\Gamma) = \int \DD h_{l}  \, Z_{BF}(\tau|h_{l}) \Psi_\Gamma(h_{l})
\label{eq:Z_BFPsi}
\ee 
where $h_{l}$ are the holonomies associated to the edges of $\Gamma = (\pp \tau)^*$, and $Z_{BF}(\tau| h_{l})$ is the natural generalization of \eqref{eq:Z_BFdiscrete} to the case of manifolds with boundaries (cf. table \ref{tab:Delta2}). 
In this generalization, all dual faces of $\Delta_3^*$ that do not intersect the boundary of $\Delta_3$ are treated as before, whereas dual faces that intersect the boundary of $\Delta_3$ are ``closed'' by links in $(\pp\Delta_3)^*$ that we also denote with $l$. 
One then associates delta functions to all dual faces, but one only integrates over the bulk holonomies $H_{l}$.
This turns the partition function into a function of the boundary variables $h_{l}$.
This function, seen as an integration kernel, allows to compute the amplitude of boundary states as in \eqref{eq:Z_BFPsi}.

It is easy to see that the amplitude kernel $Z_{BF}(\Delta_3| h_{l})$ is a projector onto flat and Lorentz invariant states $\Psi$: the boundary flatness projector is directly implemented through the bulk delta functions, whereas the boundary Lorentz invariance is implemented by group averaging.   
Then, if $\Psi$ is already Lorentz-invariant, it is easy to see that\footnote{It turns out that one of the 4 delta functions in the product above -- which would be redundant -- is actually missing, making the above formula well-defined. 
}
\be
Z_{BF}(\Psi_\Gamma) = \int \DD h_{l}  \,\prod_{f\in\Gamma} \delta\left( {\overleftarrow\prod}_{l:l\in\pp f} h_{l}^{\epsilon(l|f)}\right) \Psi_\Gamma(h_{l}) = \Psi_\Gamma(h_{l} = \bb 1).
\ee
Note that to obtain these expressions we have assumed that the topology of $\Delta_3$ is trivial\footnote{i.e. that $\Delta_3$ has the homology of the 3-ball.} as it is the case for $\tau$.
Notice  also that although this formula looks similar to \eqref{eq:Z_BFdiscrete}, there are two important differences: (\textit{i}) the presence of boundary wave function $\Psi_\Gamma$, (\textit{ii}) all quantities live now on the boundary of $\Delta_3$. 

\begin{table}[t]
\begin{center}
\begin{tabular}{|c|c|c|c|}
\hline
cell dim     & 0       &1      &2             \\ \hline
$\Delta_2 = \pp \tau$                   & vertex $v$            & edge $e$              & triangle $t$    \\ \hline
\hline
$\Delta^*_2 = \Gamma$              & face $f$          & link $l$             & node $n$              \\ \hline
cell codim                     & 2                     & 1                     & 0                           \\ \hline
\end{tabular}
\caption{Notation employed to denote cells of the two-dimensional simplicial complex $\Delta_2 = \pp \tau$ and its Poincar\'e dual $\Delta_2^* = \Gamma$. }\label{tab:Delta2}
\end{center}
\end{table}

It is now a simple observation, that by choosing $\Psi_\Gamma$ to be a spin-network state labelled by 6 spins $j_e$, this formula reproduces the Ponzano-Regge amplitude for the tetrahedron. Indeed, any boundary wave functional on $\Gamma$ that satisfies the Gauss constraint can be expanded on spin-network states, obtaining -- through steps similar to those described above -- the amplitude
\be
Z_{BF}(\Psi_\Gamma) =   \sum_{\{j_e\}} \mu_{j_e}\{ 6 j_e \} \tilde\Psi_\Gamma(j_e)
\label{eq:BFj}
\ee
with $\tilde\Psi_\Gamma(j_e)$ the components of $\Psi_\Gamma$ in the spin-network basis and $\mu_{j_e}$ encodes is the integration measure on the space of allowed spins. Whereas the (bulk) measure in \eqref{eq:Z_PR} is fixed by the demand of triangulation invariance, the measure in this formula is a priori ambiguous and depends on normalization conventions for the states. If these are taken to be normalized, then $\mu_{j_e}$ is
\be
\mu_{j_e} = \prod_{e\in\pp\tau} \sqrt{d_{j_e}} .
\ee

The goal of the rest of this article is precisely to reproduce the analogue of this derivation in the higher gauge-theoretical context, hence relating a BFCG-Yetter amplitude with boundary to the KBF model evaluated on a boundary state.

\section{BFCG theory, the Yetter model, and the KBF model \label{sec:BFCGYKBF}}
In this introductory section, we briefly review BFCG theory for the Poincar\'e 2-group, the corresponding Yetter model, as well as the KBF state sum model for a quantum flat four-dimensional spacetime.

\subsection{BFCG theory for the Poincar\'e 2-group}
Let us start by succinctly introducing (strict Lie) 2-groups from the perspective of (Lie) crossed-modules \cite{baez-lauda1}.
Let $G$ and $H$ be two groups equipped with a pair of group homomorphisms $t$ and $\alpha$ such that: $t:H\to G$ and $\alpha:G\to \mathrm{Aut}(H), g \mapsto \alpha(g) := (h\mapsto g \act h)$, i.e. an action of $G$ on $H$ by automorphisms. These homomorphisms must be compatible in the following sense: for all $g\in G$ and $h,h' \in H$, 
\be
t(g\act h) = gt(h)g^{-1} 
\quad\text{and}\quad
 t(h)\act h' = hh'h^{-1}.
\ee

The 4-dimensional Poincar\'e 2-group that we will use in the following is a simple example of this structure, where
\begin{itemize}
\item  $G=\SO(4)$, $H = \mathbb R^4$;
\item the homomorphism $t$ is trivial, $t(h)=\mathbb 1_G$ for all $h\in H$;
\item  $\alpha$ encodes the natural action of $\SO(4)$ on $\bb R^4$.
\end{itemize}
Since we will deal only with the Poincar\'e 2-group, we henceforth restrict all our formulas to this case only\footnote{The BFCG theory can be constructed for more general 2-groups \cite{Martins:2010ry}.}.

From the above definitions, one recognizes that the Poincar\'e 2-group is a different algebraic interpretation   of the standard Poincar\'e group structure. 
However, this change in perspective is absolutely crucial, since it allows to associate to the Poincar\'e group a higher topological field theory, called BFCG \cite{Girelli:2007tt}. We will argue that such higher topological field theory is better adapted to  describe discretization of 4-manifolds, than a $\SO(4)$ or a $\mathcal P=\SO(4)\ltimes \bb R^4$ BF theory. We shall comment on this point further in section \ref{sec:bfcg-BF}.

\subsubsection{Action principle \label{sec:BFCGPoincare}}

For a 2-group with trivial $t$-map ($t(h)=\bb 1_G$), the BFCG action on a 4-manifold $M_4$ is given by \cite{Girelli:2007tt}
\be
S_{BFCG} = \int_{M_4} \tr_{\mathfrak g}( B \wedge F[A]) + \tr_{\mathfrak h}(C \wedge G[\Sigma, A] ),
\ee
where the trace symbols have to be understood as appropriate $G$-invariant inner products in $\mathfrak g$ and $\mathfrak h$, respectively.  
In the Poincar\'e BFCG theory, this becomes
\be\label{eq:BFCG13}
S_{BFCG} = \int_{M_4} \tfrac12 B_{ab} \wedge F^{ab}[A]+ C_a \wedge G^a[\Sigma, A].
\ee
where $B=\tfrac12 B^{ab} J_{ab}$ and $A = \tfrac12 A^{ab} J_{ab}$ are $\so(4)$-valued 2- and 1-forms respectively, with
\be
F^{ab}[A] := \d A^{ab} + A^a{}_{c} \wedge A^{cb}
\ee
the curvature of $A$ and\footnote{Alternatively, one could have taken $(J_{ab})_{cd} = \epsilon_{abcd}$. The two choices are related by dualization of the rotation plane in $\bb R^4$.} 
\be
(J_{ab})_{cd} = 2\delta^{[a}_{c}\delta^{b]}_{d}
\label{eq:so4generator}
\ee
in the fundamental representation of $\SO(4)$.
The normalized trace symbol is thus $- \tfrac18$ of the matricial trace in the fundamental representation of $\so(4)$, i.e. 
\be
\tr_{\so(4)}(J^{ab} J^{a'b'}) := \tfrac12 \delta^{[a}_{a'} \delta^{b]}_{b'}.
\ee 
Summarizing, in the case of the Poincar\'e 2-group, the first contribution ends up being that of a 4-dimensional $\SO(4)$ BF theory.\footnote{For more general 2-gauge groups \cite{Martins:2010ry}, the definition of $F$ must be modified by a contribution involving $\Sigma$ through the differential $\tau$ of the homomorphism $t$: $F[A] \leadsto F[A,\Sigma] = \d A + \tfrac12 [A,A] + \tau(\Sigma)$. The quantity $F[A,\Sigma]$ is called the ``fake curvature''. Notice that this modification makes the two contributions to the BFCG action more symmetric.} 

The second contribution features instead a $\bb R^4$-valued 2-form $\Sigma=\Sigma^a \tau_a$, and a $\bb R^4$-valued 1-form $C=C^a \tau_a$  where $ \tau_a$ is an orthonormal basis of $\bb R^4$ as a vector space. 
The trace $\tr_{\bb R^4}$ taken to be the canonical $\SO(4)$-invariant inner product
\be
\tr_{\bb R^4}( \tau_a  \tau_b) = \delta_{ab}.
\ee
The quantity $G$, defined as
\be
G^a[\Sigma,A]:=(\d \Sigma + A \act \Sigma)^a \equiv \d \Sigma^a + A^a{}_b \wedge \Sigma^b,
\label{eq:2-curvG}
\ee 
is a 3-form and represents the {\it 2-curvature} of $\Sigma$. In fact, $\Sigma$ is best understood as a 2-connection. To see this  we need to discuss the transformation properties of the fields $A$, $B$, $C$, and $\Sigma$. 

Before delving into this, let us consider the equations of motion of BFCG theory. These are
\begin{subequations}
\begin{alignat}{5}
&&\text{(1-flatness) $\delta B$: } &F[A] =0\qquad
&\text{(2-flatness) $\delta C$: } &G[\Sigma,A] =0\\
&&\text{  $\delta A$: } &\d_A B - 2C\wedge \Sigma = 0\qquad
&\text{(torsion freeness) $\delta \Sigma$: } & T[C,A]  = 0.
\end{alignat}
\end{subequations}
where the ``torsion'' $T$ is defined as (this naming will be clarified in the forthcoming sections):
\be
T[C,A] := \d_A C \equiv \d C + A\act C = ( \d C^a + A^a{}_b \wedge C^b) \tau_a.
\ee

When convenient, we will identify forms valued in $\bb R^4\wedge \bb R^4$ with forms valued in $\so(4)$ according to e.g.
\be
C\wedge\Sigma \equiv  \tfrac12 C^{[a}\wedge \Sigma^{b]}  J_{ab}.
\ee

\subsubsection{Symmetries and constraints of BFCG theory }\label{sec:diffeos}

In the Poincar\'e case, the BFCG action is invariant under the following action of a 1- and 2-Lorentz gauge transformations:
\begin{subequations}
\be
\text{1-Lorentz: } 
\begin{cases}
A \mapsto A + \d_A \xi\\
B \mapsto B + [B, \xi]\\
C \mapsto C - \xi \act C\\
\Sigma \mapsto \Sigma - \xi \act \Sigma\\
\end{cases} 
\quad\text{and}\quad
\text{2-Lorentz: } 
\begin{cases}
A \mapsto A \\
B \mapsto B - 2C \wedge \eta \\
C \mapsto C \\
\Sigma \mapsto \Sigma + \d_A \eta \\
\end{cases}
\label{eq:12Lorentz}
\ee
for infinitesimal $\so(4)$-valued 0-forms $\xi$, and $\bb R^4$-valued 1-forms $\eta$.
These are complemented by the 1- and 2-shift gauge symmetries 
\be
\text{1-shift: }
\begin{cases}
A \mapsto A \\
B \mapsto B +  \d_A \lambda\\
C \mapsto C \\
\Sigma \mapsto \Sigma \\
\end{cases}
\quad\text{and}\quad
\text{2-shift: }
\begin{cases}
A \mapsto A \\
B \mapsto B + 2\mu\wedge\Sigma \\
C \mapsto C + \d_A \mu \\
\Sigma \mapsto \Sigma \\
\end{cases}
\label{eq:12shift}
\ee
\end{subequations}
{ for infinitesimal $\so(4)$-valued 1-form $\lambda$, and $\bb R^4$-valued 0-form $\mu$.}

As in the BF  case, the  1-shift symmetry is reducible. The 2-shift symmetry is however not reducible.  

In BFCG theory, diffeomorophism symmetry can be expressed -- on-shell of the equations of motion -- in terms of field-dependent internal 1- and 2-gauge symmetries of the fundamental fields \cite{Belov:2018uko}. 
In particular, the action of an infinitesimal diffeomorphism $X\in\mathfrak{X}^1(M)$ translates into the combined action of the  following  internal symmetries:
\be
\Big(\xi_X = \iota_X A, \quad \lambda_X = \iota_X B, \quad \eta_X = \iota_X \Sigma, \quad \mu_X = \iota_X C \Big)
\ee 
i.e.
\be
\text{diffeos: }
\begin{cases}
A \mapsto A + \pounds_X A = A + \iota_X F + \d_A \xi_X \\
B \mapsto B + \pounds_X B = B + [B,\xi_X] + \d_A \lambda_X - 2C\wedge \eta_X + 2\mu_X \wedge\Sigma \\
C \mapsto C + \pounds_X C = C + \iota_XT -\xi_X \act C + \d_A \mu_X  \\
\Sigma \mapsto \Sigma + \pounds_X \Sigma = \Sigma - \xi_X \act \Sigma + \d_A \eta_X \\
\end{cases}
\ee 
These formulas are easily proved using Cartan's formula $\pounds_X = \iota_X \d + \d \iota_X$.
This result is in complete analogy to what happens in $BF$ theory (in any dimensions), where diffeomorphisms can also be expressed, on-shell of the equations of motion, as a combination of (field-dependent) internal symmetries,  as we explained in section \ref{sec:BF2PR}. 

Moreover, it is not hard to check that the only solutions to the equations of motions are given by pure gauge configurations, a fact that indicates that the theory is topological.

Although we do not provide a complete Hamiltonian analysis of the theory (see for example \cite{Mikovic:2015hza}), we still mention some relevant facts. 
Off-shell of the constraints, i.e. at the kinematical level, the phase space variables (excluding Lagrange multipliers) are obtained via pullback of the fundamental fields onto a Cauchy hypersurface, $\{a=\iota^*A, b=\iota^*B, c=\iota^*C, \sigma=\iota^*\Sigma\}$, whereas the Poisson brackets pair $a$ to $b$ and $c$ to $\sigma$. Then, the phase space analogue of the above symmetries is Poisson-generated by the following constraints, obtained via pullback of the equations of motion:
\begin{subequations}
\begin{alignat}{2}
\text{1-Gauss: } & {}^1\!\mathcal{G} := \d_a b  - 2 c\wedge \sigma \approx 0 &&\q\leadsto\q \text{1-Lorentz} \label{1-GaussC}\\
\text{2-Gauss: } & {}^2\!\mathcal{G} := T[c,a] \equiv \d_a c \approx 0 &&\q\leadsto\q \text{2-Lorentz} \label{2-GaussC}\\
\text{1-flatness: } & {}^1\!\mathcal{F} := F[a] \equiv \d a + a \wedge a \approx 0 &&\q\leadsto\q \text{1-shift} \\
\text{2-flatness: } &  {}^2\!\mathcal{F} := G[\sigma,a] \equiv \d \sigma + a \triangleright \sigma \approx 0  &&\q\leadsto\q \text{2-shift} 
\end{alignat}
\label{eq:BFCGconstraints}
\end{subequations}
The BFCG constraint algebra is then
\begin{align}
\{{}^1\!\mathcal{G}[\xi],{}^1\!\mathcal{G}[\xi']\} &= {}^1\!\mathcal{G}[[\xi,\xi']], & \{{}^1\!\mathcal{G}[\xi], {}^2\!\mathcal{F}[\mu] \} &= {}^2\!\mathcal{F}[\xi\triangleright \mu],\notag\\
\{{}^1\!\mathcal{G}[\xi], {}^2\!\mathcal{G}[\eta]\} &= {}^2\!\mathcal{G}[\xi\triangleright \eta], & \{{}^1\!\mathcal{F}[\lambda], {}^1\!\mathcal{G}[\xi]\} &= {}^1\!\mathcal{F}[[\lambda,\xi]],\notag\\
\{ {}^2\!\mathcal{F}[\mu], {}^2\!\mathcal{G}[\eta]\} &= {}^1\!\mathcal{F}[\mu\wedge \eta],
\end{align}
where we recall that the smearing parameters $\xi$ and $\mu$ are 0-forms valued in $\mathfrak{so}(4)$ and $\mathbb{R}^4$ respectively, while  $\lambda$ and  $\eta$ are 1-forms valued in $\mathfrak{so}(4)$ and $\mathbb{R}^4$, respectively.

\subsubsection{BFCG partition function}\label{sec:bfcgpart}

Formally, the partition function of the BFCG theory on a closed and compact 4-manifold $M_4$ is
\be
Z_{BFCG}(M_4) = \int \DD B \DD A \DD C \DD\Sigma\, e^{i \int_{M_4} \tr( B\wedge F[A] ) + \tr(C\wedge G[\Sigma,A])} .
\label{eq:Z_BFCG}
\ee
Integrating out the 1- and 2-curvature Lagrange multipliers $B$ and $C$, we obtain
\be
Z_{BFCG}(M_4) = \int \DD A \DD\Sigma\, \delta( F[A] ) \delta( G[\Sigma,A]) .
\label{eq:Z_BFCG_cont}
\ee

\subsection{From BFCG theory to the Yetter model \label{sec:BFCG2Y}}

We now want to discretize this formal expression on a simplicial decomposition $\Delta_4$ of $M_4$. 
To do so, we introduce smeared versions of the 1- and 2-connections $A$ and $\Sigma$ onto the links $l\in\Delta_4^*$ and dual face $f\in\Delta_4^*$ -- where $\Delta_4^*$ is the Poincar\'e-dual of $\Delta_4$ (cf. table \ref{tab:Delta4}):
\be
H_{l} = \mathrm{Pexp}\int_{l} A \in \SO(4)
\quad\text{and}\quad
X_{f}(n) = \mathrm{Sexp} \int_{f} \Sigma' \in \mathbb R^4.
\label{eq:AXdiscrete}
\ee
The presence of the {\it  face} variables $X_{f}$ is the true novelty of this higher-gauge theory approach to the Poincar\'e topological field theory.
In these formulas, $\mathrm{Sexp}$ indicates a ``surface ordered'' exponential\footnote{The surface-ordering is not really necessary since $(\bb R^4,+)$ is Abelian.} and the prime in $\Sigma'$ indicates that $\Sigma$ must be appropriately parallel-transported by $A$ and a choice of paths at a common reference point -- here taken to be a dual node (0-cell) $n \in\Delta_4^*$ -- before integration.

Hence, whereas $H_{l}$ is simply attached to an (oriented) link, $X_{f}(n)$ is attached to a pair $(f,n)$ with $n\in\pp f$.  
The operation of modifying the choice of $n\in\pp f$ is called {\it whiskering}, and the action of a certain product of $H_{l}$'s on $X_{f}(n)$, i.e. it involves a parallel transport operation along a path of links, 
\be
X_{f}(n_k) =  (H_{n_k n_{k-1}} \cdots H_{n_2 n_1}) \act X_{f}(n_1).
\label{eq:Xtransport}
\ee
We will come back to this point in the following section, and for now consider the variables $X_{f}(n)$ to be defined at certain (fixed) nodes $n$.

\begin{table}[t]
\begin{center}
\begin{tabular}{|c|c|c|c|c|c|}
\hline
cell dim     & 0       &1      &2        & 3      & 4         \\ \hline
$\Delta_{ 4}$                   & vertex $v$            & edge $e$              & triangle $t$          & tetrahedron $\tau$   &    4-simplex $\sigma$ \\ \hline
\hline
$\Delta^*_{ 4}$        &  --               & bubble $b$              & face $f$          & link $l$             & node $n$              \\ \hline
cell codim                & 4         & 3                     & 2                     & 1                     & 0                           \\ \hline
\end{tabular}
\caption{Notation employed to denote cells of the four-dimensional simplicial complex $\Delta_4$ and its Poincar\'e dual $\Delta_4^*$.}\label{tab:Delta4}
\end{center}
\end{table}

Discretizing on a simplicial decomposition $\Delta_4$ of $M_4$, formally gives the Yetter model \cite{yetter1993tqft} for the Poincar\'e 2-group\footnote{The discretization scheme proposed differs from the one of \cite{Girelli:2007tt}, 
insofar as our connection variables are smeared on the Poincar\'e dual of a simplicial complex, rather than on the simplicial complex itself.} \cite{Girelli:2007tt}: 
\begin{align}\label{eq:discreteZ1}
&Z_{BFCG}(\Delta_4) \equiv Z_{Y}(\Delta_4) = \int\!\! \DD H_{l} \DD X_{f}\!\!
\prod_{f\in\Delta_4^*}\!\!\delta_{\SO(4)}\!\!\left( \overleftarrow{\prod_{l\in\pp f}} H_{l}^{\epsilon(l|f)} \right) 
\prod_{b\in\Delta_4^*}\!\!\delta_{\bb R^4}\!\!\left( \sum_{f\in\pp b} \!\! \epsilon(f|b) X'_{f}(n) \right)
\end{align}
where $\DD H_{l} $  and $\DD X_{f} $ are products of Haar measures on $\SO(4)$ and $\mathbb R^4$ respectively, $b$ is a 3-cell of $\Delta_4^*$ and where, again, the prime in $X'_{f}(n)$ denotes the need of some parallel transport of the form \eqref{eq:Xtransport} which is left understood. This need spurs from the fact that not all the $X_{f}$ bounding a dual 3-cell can be defined at the same vertex.  Thus we need to define how to parallel transport the $X_f$-variables inside a given 3-cell $b$.

Crucially, the choice of paths needed for these parallel transport operations is irrelevant on-shell of the (discrete) 1-flatness encoded in the first delta function. 

Let us note that, like the partition function \eqref{eq:Z_BFdiscrete} of the discrete BF theory, this partition function is also only formal: it will, in general, diverge. 
However, when computed on a finite 2-group,\footnote{In this case the homomorphism $t$ might not be trivial.} it defines an actual topological invariant of the discretized 4-manifold, the Yetter invariant  \cite{yetter1993tqft}.

\subsection{The KBF model\label{sec:KBF}}

Let us finally describe the KBF model. It shares many ingredients with Regge calculus. However, it is topological, and as such not directly related to discrete general relativity.

Korepanov constructed the amplitudes of the KBF model \cite{Korepanov:2002tr,Korepanov:2002tq,Korepanov:2002tp}, based on some interesting invariance relations for certain geometric quantities under Pachner moves. 
Baratin and Freidel \cite{Baratin:2006gy} provided a definition of the state sum for the model, together with an in-depth analysis of the intricate, non-compact, gauge symmetries. 
This in turn allowed them to devise a Faddeev-Popov procedure to gauge fix the redundant non-compact symmetries, hence providing a definition of the state sum that is free of divergences as well as a rigorous proof of the (bulk) triangulation invariance of the KBF model. 

Moreover, Baratin and Freidel also showed that the KBF model can be formulated as a 2-categorical state sum model associated to a simplicial decomposition of a 4-manifold \cite{Baratin:2009za,Baratin:2014era}.
This means that the under its geometric surface, the KBF model hides a tight relation with the (higher!) representation theory of the Poincar\'e 2-group \cite{Baez:2008hz,Baratin:2014era}.

The KBF state sum model is defined as follows.
Given a simplicial decomposition $\Delta_4$ of a 4-manifold, label (cf. table \ref{tab:Delta4})
\begin{itemize}
\item its edges $e\in\Delta_4$ with a positive real number $l_e \in \mathbb R_+$ representing its length, and
\item its triangles an integer number $s_t \in\mathbb Z$;\end{itemize}
then the state sum model is obtained by summing over all possible labels a product of local weights associated to the various simplicial cells, as follows:
\begin{subequations}
\be
Z_{KBF}(\Delta_4) = \int \DD l_{e} \DD s_t \, 
\prod_{t\in\Delta_4} 2 A_t(\{l_e\}) 
\prod_{\tau\in\Delta_4} (-1)^{\sum_{t:t\in\tau}s_t}
\prod_{\sigma\in\Delta_4} \frac{\cos\Big( S_{KBF} \Big) }{4! \mathrm{Vol}_\sigma( \{l_e\} )} \label{eq:Z_KBF0}
\ee
with 
\be
S_{KBF} = \sum_{t:t\in\sigma} s_t \Theta_t(\{l_e\})
\ee
\label{eq:Z_KBF}
\end{subequations}
where
\begin{itemize}
\item $\DD \ell_e$ is the product of the Lebesgue measures $\d l^2_e$,
\item $\DD s_t$ is the product of the the counting measure $\tfrac{1}{2\pi}\sum_{s_t}$,
\item $A_t$ is the area of the triangle $t\in\Delta_4$, 
\item $\mathrm{Vol}_\sigma$ is the 4-volume of the four simplex $\sigma\in\Delta_4$, and 
\item $\Theta_t$ is the exterior (hyper-)dihedral angle between the two tetrahedra that meet at the triangle $t$.
\end{itemize}
{\it All} the geometric quantities above are defined as functions of the side lengths $\{l_e\}$, which fully determine the shape of a four simplex $\sigma$.

The measure factor is fixed by the requirement of triangulation invariance of the model \cite{Korepanov:2002tp, Korepanov:2002tq, Korepanov:2002tr, Baratin:2009za,Baratin:2014era}.

We conclude this section with a comparison of the KBF quantum partition function and (quantum) Regge calculus.
In particular, we notice that the KBF action $S_{KBF}$ \eqref{eq:Z_KBF} differs from the 4-dimensional Regge action $S_R^{d=4}$ \eqref{eq:Regge_d}: the label $s_t$ -- morally occupying the place of the triangle areas -- is assigned independently from the edge lengths $l_e$, and therefore is generically non-geometric. The variables $s_t$ act as Lagrange multipliers and impose flatness. 
Notice also that the relation with the Regge action, albeit incomplete, is more direct than in three dimensions, as there is \textit{no} semiclassical (``large spin'') limit necessary to reveal it. 
Finally, the appearance of a cosine -- as opposed to an imaginary exponential -- has the natural interpretation of encoding a sum over orientations, in complete analogy to the Ponzano-Regge model \eqref{eq:Regge_asympt} \cite{Christodoulou:2012af}.


\section{Tetrad variables and edge simplicity in the BFCG-Yetter model\label{sec:4}}

In the previous section we have reviewed BFCG-Yetter theory and the KBF model. 
In the rest of the paper, we will develop  tools to prove the correspondence between the two. 
This section focuses on one of the main technical ingredients of this work: ``edge simplicity''.

We will start by reinterpreting the $C$ field of BFCG theory as a tetrad field, $C\equiv e$.
This will allow us to note how, on-shell of the 2-Gauss constraint, the BFCG action bears strong similarities to the KBF action $S_{KBF}$.
We will then discuss the classical continuum origin of the edge simplicity and the 2-Gauss constraints, and how edge simplicity is subsumed by, but weaker than, 1-flatness.

We will then move back to the discrete realm, where first we identify the role that edge simplicity plays when interpreting the BFCG-Yetter partition function geometrically, and then discuss its interplay with the discrete version of 1-flatness.   
These steps will greatly benefit from us focusing the study of the BFCG-Yetter model on a single four-simplex and, in particular, on its boundary.

\subsection{Tetrad field and torsionfreeness in the continuum\label{sec:contedgesimplicity}}

One of the main difficulties that 4-dimensional spinfoam models based on a 4-dimensional BF theory need to face is that they need to extract a (unique) tetrad field $e$, hence the associated Levi-Civita connection, from the fundamental {\it bivector} field $B$. 
This difficulty percolates down to the coupling of matter to gravity and in particular to how to couple torsion degrees of freedom (fermions) to the spinfoam model.

An advantage of the BFCG model, is that it provides, in addition to the bivector variables, the actual tetrad variables. 
Indeed, in the Poincar\'e BFCG theory the field $C$ is a $\bb R^4$-valued 1-form and can be directly related with the {\it tetrad} field $E_\mu^a$:
\be
C \equiv E.
\ee
With this interpretation of the $C$ field, one readily sees -- via integration by parts of the $C\wedge G\equiv E \wedge\d_A \Sigma$ term in the BFCG action \eqref{eq:BFCG13} -- that this explicitly features not only the curvature $F$ of $A$, but also its torsion $\d_A E$ with respect to the tetrad field $E$ \cite{Mikovic:2015hza}:
\be
\label{eq:BFCG33}
S_{BFCG} = \int_{M_4} \tr_{\so(4)}( B\wedge F[A]) + \tr_{\bb R^4}( \Sigma \wedge \d_A E ) + \d \tr_{\bb R^4}(\Sigma \wedge E),\qquad \q(C\equiv E).
\ee
With the BFCG action put in this form, it is also manifest that the 2-connection $\Sigma$ is nothing else than a Lagrange multiplier enforcing the torsionless condition of $A$ as much as $B$ is a Lagrange multiplier for its flatness: in absence of torsion, 
\be
T[A,E] = \d_A E \approx 0,
\label{eq:torsionless}
\ee
one has $A = A_{LC}(E)$ (Levi-Civita) and,  neglecting the boundary term, the BFCG action reduces to
\be
S_{BFCG} \, \stackrel{\d_A E = 0}{\approx}\, \int_{M_4} \tr_{\so(4)}( B\wedge R[E]) ,
\label{eq:BFCG_torsionless}
\ee
where we defined the Riemann curvature 2-form $R[E] := F[A_{LC}(E)]$.

In equation \eqref{eq:BFCG_torsionless}, we recognize the continuum analogue of the KBF action \eqref{eq:Z_KBF}: the Riemann curvature corresponds to deficit angles expressed a function of the edge lengths (discrete metric), whereas the generic bivector field $B$ corresponds to the independent variable $s_t$.

In the next section, we will analyze the discrete quantum model of BFCG theory. 
There, we will want to impose the torsionless, or 2-Gauss, constraint on the boundary states. 
Since torsion is a 2-form, the torsionless condition naturally translates into a condition, the ``closure condition'', to be imposed at boundary triangles.
The problem is that, as we will show, triangle closures are a priori not well defined {\it throughout} the whole discretization.
This is because one needs to parallel transport one given discrete edge variable to different tetrahedron frames to feature in different triangle closures.

Imposing the 1-flatness constraint would clearly solve this issue, but would at the same time completely kill all the connection degrees of freedom (after all we want to keep at least $A=A_{LC}(E)$ above). 
The solution to this problem consists in realizing that, in order to fully impose the 2-Gauss/torsionless constraint at the discrete level, imposing triangle closures is not enough: edge simplicity is also required. 

Although trivial in the continuum, the covariant derivative of the 2-Gauss/torsionless constraint \eqref{eq:torsionless}, i.e. 
\be
\text{continuum edge simplicity:}\quad \d_a(\d_ae)=F[a]\triangleright e \approx 0,
\label{eq:contedgesimplicity}
\ee
(as usual $e=\iota^*E$),
provides {\it in the discrete picture} the condition -- independent from the triangle closures -- that makes the latter consistent throughout the discretization.
In fact, in the next section we will discuss how, in the discrete picture, edge simplicity renders the very notion of edge vector well defined.  

Notice that, although we derived it from the 2-Gauss/torsionless constraint, continuum edge simplicity is automatically satisfied for any flat connection.
An important part of the present work will consist in explicitly splitting the (discrete) flatness constraint of the Yetter model into edge simplicity -- reinterpreted as part of the 2-Gauss constraint -- and a remaining condition. This remaining condition will be shown to constrain the extrinsic geometry of the four-simplex' boundary to be the one associated to an embedding into flat Euclidean space (section \ref{sec:1flatness4LC}).

\subsection{Bulk edge vectors, triangle closures, and edge simplicity}

The Yetter model partition function \eqref{eq:discreteZ1} is expressed in the $(H_l, X_f)$ polarization, corresponding to the continuum pure-connection polarization $(A, \Sigma)$ \eqref{eq:Z_BFCG_cont}. 
The delta functions make explicit the fact that one is integrating over 1- and 2-flat configurations.
However, the invariance of both the measure and the integrand of \eqref{eq:discreteZ1} under 1- and 2-gauge transformations\footnote{From the formulas we provided, this can be easily checked in the continuum. In the discrete, the same holds true, even though we restrain from providing explicit expressions for the off-shell action of these symmetries in the discrete bulk for we will not need them. However, we will need and thus provide explicit formulas for the action of the symmetries on the corresponding boundary variables.} implies that \eqref{eq:discreteZ1} is also invariant under these symmetries -- which are thus imposed through a group-averaging mechanism in analogy with what we discussed for 3d BF theory. 

To make contact with the discussion of the previous section, and thus move the emphasis on the tetrad field $C\equiv E$ and torsionfreeness rather than 2-flatness, one can perform a Fourier transform on the discrete 2-holonomies $X_f\in\mathbb R^4$ to dual variables $L_e\in\mathbb R^4$, $e\in\Delta_4$ \cite{Mikovic:2011si}.
This dualization is suggested by applying a Fourier transform to the 2-flatness conditions (recall that $e=b^*$ and vice versa, thus $L_e \equiv L_b$) 
\be
\delta_{\bb R^4}\!\!\left( \sum_{f:f\in\pp b} \!\! \epsilon(f|b) X'_{f}(n) \right)\,=\,\frac{1}{(2\pi)^4} \int_{\mathbb{R}^4} \!\!\d^4 L_{b} \exp\left(i \, L_{b}\cdot \! \!\! \sum_{f:f\in\pp b}  \epsilon(f|b)X'_{f}  \right) ,
\label{eq:delta2Flat}
\ee
and the subsequent re-expression of the BFCG-Yetter partition function \eqref{eq:discreteZ1} solely in terms of the $L_b$ variables, after integration of the $X_f$:
\begin{align}
Z_{Y}(\Delta_4)=\int \DD H_{l} \DD L_{b}\!
\prod_{f\in\Delta_4^*}\delta_{\SO(4)}\!\!\left( \overleftarrow{\prod_{l:l\in\pp f}} H_{l}^{\epsilon(l|f)} \right) 
\prod_{f\in\Delta_4^*}\delta_{\bb R^4}\!\!\left( \sum_{ b: f\in\pp b }  \!\!\epsilon(f|b) L'_{b} \right).
\label{eq:discreteZ2}
\end{align}
Notice that $L_b\equiv L_e$ is a four-vector associated to an edge, and that -- by dualizing the labels -- the second delta function in \eqref{eq:discreteZ2} implements nothing else than the triangle closure, 
\be
\sum_{ e: e \in t }  \!\!\epsilon(e|t) L'_{e} = 0 . \label{4-2-gauss}
\ee
As discussed in the previous section, this corresponds to the torsionless condition $\d_A E =0$ discretized over triangles as soon as we identify the edge vector with the smeared tetrad field, 
\be
L_e = \int_e E.
\ee
Also, we stress that \eqref{4-2-gauss} hides in the primed notation the presence of $H$-parallel transports of the $L$ variables. 

Notice that the Fourier transform performed here between the $X$ and $L$ variables parallels the integration by parts used to put the BFCG action in the form \eqref{eq:BFCG33}. In particular, both procedures make explicit the torsionless condition. In the discrete picture this is simply the set of triangle closures.

Two remarks are in order at this point. First, the presence of these parallel transports, and their involved dependence on the variables $H_f$, is what prevents one to successfully apply the $\SO(4)$ recoupling theory to \eqref{eq:discreteZ2} and obtain a state sum amplitude in analogy with the first derivation of the Ponzano-Regge model in section \ref{sec:BF2PR}. 
This is not particularly bad news, though: for the use of $\SO(4)$ recoupling theory would lead to $\SO(4)$ representation labels for the corresponding state sum model, which profoundly differ from the representation labels of the Poincar\'e 2-group featured in KBF.\footnote{We also notice that both these sets of labels differ from the representation labels of the Poincar\'e group $\mathcal P=\SO(4)\ltimes \bb R^4$ which would appear if we were discretizing the BFCG action as we would discretize a standard BF theory with the standard Poincar\'e group $\mathcal P$ as a gauge group (cf. discussion at the end of section \ref{sec:bfcg-BF}).}

Second, and even more importantly, it turns out that the triangle closure constraints heavily depend on the parallel transport conventions one chooses. 
Of course, on-shell of 1-flatness, this dependence on conventions disappears.
However, we want to interpret triangle closures as torsionless conditions to be implemented independently of 1-flatness (see the discussion around \eqref{eq:BFCG_torsionless}).
As anticipated in the previous section, the source of the problem is that at the discrete level torsionfreeness is not fully captured by triangle closure conditions: the latter must be complemented with edge simplicity conditions.
These ensure that an edge vector is stabilized by the holonomy around it, which in turn means that the edge vector can be consistently parallel transport to all the  tetrahedron reference frame around it -- a procedure necessary to write down all the triangle closures. 

Incidentally, as the discussion around equation \eqref{eq:torsionless} shows, imposing the edge simplicity constraints also provides a definition of discrete Levi-Civita connection.
 Notice that this observation strongly suggests that edge simplicity is a restriction on the 1-connection $H_l$ rather than on the edge vectors\footnote{ To the reader accustomed to LQG this result might be surprising, since the closely related LQG triangle closure conditions  give restrictions on the fluxes rather than on the connection.} $E_e$. 
It is then important to identify the remaining freedom present in $H_l$ to properly write down the geometric 1-flatness condition and thus impose it in terms of $\delta$-functions over unconstrained variables. 
This step is crucial because Fourier transforming  these last $\delta$-functions we will find the phase factors that define the KBF model.

The execution of this program is somewhat easier if we focus on the amplitude of a single 4-simplex. 
The full partition function can then be re-obtained by ``gluing'' the various 4-simplex amplitudes, i.e. by appropriately identifying and integrating over the boundary data of neighbouring 4-simplices.

\subsection{BFCG-Yetter model with boundaries}\label{sec:BFCG-Yetter}

We restrict now our attention to a single four-simplex $\sigma$, and its boundary 3-complex $\Gamma = (\pp \sigma)^*$, see table \ref{tab:Gamma}.
Generalizing what happens in 3d BF theory with boundaries, we notice that the boundary link $l$ and the boundary $f$ ``close'' the bulk face $f(l)\in\sigma^*$ and the bulk bubble $b(f)\in\sigma^*$ which intersect the boundary of $\sigma$.
Thus, to the links and faces of $\Gamma$, we attach the variables $h_l$ and $x_f$, respectively.

\begin{table}[t]
\begin{center}
\begin{tabular}{|c|c|c|c|c|}
\hline
cell dim     & 0       &1      &2        & 3               \\ \hline
$\Gamma^* = \pp \sigma = \Delta_3$                   & vertex $v$            & edge $e$              & triangle $t$          & tetrahedron $\tau$    \\ \hline
\hline
$\Gamma = (\pp \sigma)^*=\Delta_3^*$                 & bubble $b$              & face $f$          & link $l$             & node $n$              \\ \hline
cell codim                   & 3                     & 2                     & 1                     & 0                           \\ \hline
\end{tabular}
\caption{We recall the notation employed to denoted cells in a 3-complex, table \ref{tab:Delta3}, while specializing to $\Gamma$, the 3-complex dual to the boundary of a 4-simplex $\sigma$. 
}\label{tab:Gamma}
\end{center}
\end{table}

These variables allow us to define the amplitude kernel $Z_Y (\sigma | h_{l}, x_{f})$: it is the BFCG-Yetter amplitude of $\sigma$ where the dual bulk faces and bubbles that intersect the boundary (all of them in the 4-simplex case) are completed with the boundary data we just introduced, i.e. $h_l$ and $x_f$ respectively -- over which we do {\it not} integrate\footnote{For simplicity, we have assumed that the boundary link $l$ has the same orientation as the corresponding bulk face $f(l)$ (i.e. the bulk face such that $l=f(l)\cap \Gamma$, and similarly for the boundary face $f$ and bulk bubble $b(f)$.}.
\begin{align}
Z(\sigma| h_l, x_f) = \int \DD H_l \DD X_f \; 
&\prod_{f:f\in\sigma^*} \delta_{\SO(4)}\Big( h_{l=f\cap \Gamma} {\overleftarrow\prod}_{l:l\in f \cap \mathring{\sigma}^*} H^{\epsilon(l|f)}_l\Big)\times\notag\\
&\times\prod_{b:b\in\sigma^*} \delta_{\bb R^4} \Big( x'_{f=b\cap\Gamma}+ {\sum}_{f:f\in b\cap\mathring{\sigma}^*} \epsilon(f|b)X'_f \Big)
\end{align}
Finally, integrating this amplitude kernel against a boundary wave-function $\Psi_\Gamma(h_{l},x_{f})$, we obtained the boundary amplitude  
\be
Z_Y(\sigma|\Psi_\Gamma) = \int \DD h_{l} \DD x_{f} \, Z_Y (\sigma | h_{l}, x_{f}) \Psi_\Gamma(h_{l},x_{f}) \, ,
\label{eq:BFCGYkernel}
\ee
see section \ref{app:1} for the details.
In analogy with the 3d BF result \eqref{eq:discreteZ2}, one finds that integrating out the bulk data $H_l$ and $X_f$, the bulk amplitude imposes the 1- and 2-flatness constraints on the boundary 1- and 2-connections:
\begin{align}
\label{eq:discreteZ3}
Z_Y(\sigma|\Psi_\Gamma) 
=\int \DD h_{l} \DD x_{f}   \prod_{f\in\Gamma}\delta_{\SO(4)} \bigg( \overleftarrow{\prod_{l\in\pp f}} h_{l}^{\epsilon(l|f)} \bigg)   
 \prod_{b\in\Gamma}\delta_{\bb R^4}\!\bigg( \sum_{f\in\pp b} \!\! \epsilon(f|b) x'_{f}\bigg)
 \Psi_\Gamma(h_{l},x_{f}) .
\end{align} 

Before proceeding with the dualization of the 2-connection data, a few comments are in order:
\begin{itemize}
\item In general, some of the bulk delta functions defining the kernel \eqref{eq:BFCGYkernel} are redundant, and integrating out the bulk variables produces spurious divergences. In analogy with 3d BF theory \cite{Freidel:2004vi, Barrett:2008wh, Bonzom:2010ar, Bonzom:2010zh}, these divergences also correspond to the bulk 1- and 2-shift symmetries and should be gauge-fixed through a Faddeev-Popov procedure. This is detailed for the KBF model in \cite{Baratin:2006gy}. In the present case of a single 4-simplex, as shown in appendix \ref{app:1}, no redundant delta functions appear (this is because the bulk triangulation contains no bulk vertices and no bulk edges) and the answer is automatically finite. To keep the notation lighter, in the following we will not explicitly indicate which delta functions appear in the amplitude.
\item In integrating out the bulk variables, and solving for the bulk delta functions, one actually obtains only 6 boundary-face delta functions on the $h_l$ -- instead of 10, -- and only 4 boundary-bubble delta functions on the $x_f$ -- instead of 5, see appendix \ref{app:1}. These are enough to deduce the complete 1- and 2-flatness of the boundary data. In equation \eqref{eq:discreteZ3}, we leave this fact understood.
\item Whereas the boundary data of a 4d BF or LQG amplitude is labelled by a 2-complex, in the BFCG-Yetter model, the boundary data is labelled by the (dual) \textit{3}-complex $\Gamma$. Dualizing, we see that the 3-complex contains enough information to reconstruct not only the edges of the boundary triangulation (dual to faces of $\Gamma$), but also its vertices (dual to the bubbles of $\Gamma$). As we will see in section \ref{sec:constraints},  the boundary 1- and 2-flatness constraints, encoded in the boundary delta functions of \eqref{eq:discreteZ3} generate translations of the boundary edges and vertices, respectively. This is a crucial advantage of this formulation, since it makes a full implementation of (discrete) diffeomorphism symmetry possible. 
\item In \eqref{eq:discreteZ3}, the boundary 1- and 2-flatness constraints are explicitly imposed via delta functions. They correspond to the last two equations of  \eqref{eq:BFCGconstraints}. In complete analogy with the 3d BF case, equation \eqref{eq:discreteZ3}  imposes the 1- and 2-Gauss constraints (i.e. the first two equations of  \eqref{eq:BFCGconstraints}) on $\Psi$ via group averaging. This is a consequence of the invariance of both the integration kernel and the integration measures under 1- and 2-Lorentz transformations of the boundary variables. 
\end{itemize}

This last point can be made manifest through a Fourier-transformation of the $\bb R^4$ delta functions.
This mimics what done in \eqref{eq:discreteZ2}, with the difference that now only the boundary data is involved.
Thus, writing
\be
\label{eq:delta27}
\delta_{\bb R^4}\!\bigg( \sum_{f\in\pp b} \!\! \epsilon(f|b) x'_{f}\bigg)=\frac{1}{(2\pi)^4}
\int_{\mathbb{R}^4} \d^4 v_{b} \;\exp\bigg(i \, v_{b}\cdot\! \! \!\! \sum_{f\in\pp b}  \epsilon(f|b)x'_{f}  \bigg) ,
\ee
and defining the Fourier transformed $\tilde\Psi$ through
\be
\label{eq:Ftrafo0}
 \Psi_\Gamma(h_{l},x_{f}) =: \int \DD \ell_{f} \, e^{ i \sum_{f} \ell_{f} \cdot x_{f}. }\tilde \Psi_\Gamma(h_{l} , \ell_{f} ),
\ee
we can now integrate out the $x_f$ variables explicitly from \eqref{eq:discreteZ3}, and hence obtain -- instead of the 2-flatness constraints -- the following product of delta functions:
\be
\label{eq:delta29}
\prod_{f\in\Gamma} \delta_{\bb R^4} \bigg( \ell_{f}  - (v'_{b=t(f)} - v'_{b=s(f)} )   \bigg) 
\equiv
\prod_{e\in\pp\sigma} \delta_{\bb R^4} \bigg( \ell_{e}  - (v'_{v=t(e)} - v'_{v=s(e)} )   \bigg). 
\ee  
Primes, as usual by this point, denote the need of appropriate parallel transportations of the primed variables through the action of a combination of $h_l$'s. Here, they descend from the primes in \eqref{eq:delta27}, through the $\SO(4)$ invariance of the $\bb R^4$ inner product: $v\cdot (h\triangleright x) = (h^{-1}\triangleright v)\cdot x$.
On the right-hand-side of \eqref{eq:delta29}, we have simply dualized all sets of labels from denoting cells in $\Gamma=(\pp \sigma)^*$ to denoting cells in $\pp \sigma$.
The dualization of the labels allows us to easily decode the geometrical meaning of the above expressions (at least on-shell of 1-flatness where the primes can be safely ignored), and see that the $v_v\in\bb R^4$ denotes the position of the vertex $v$ of the four-simplex $\sigma$ within $\bb R^4$, while $\ell_e\in\bb R^4$ denotes the corresponding edge vector. 

As a last step towards the rewriting of the BFCG-Yetter boundary amplitude in a way that makes the 2-Gauss constraints manifest, we integrate out the $v$ variables which, at this point, only appear inside the delta functions \eqref{eq:delta29}. 
This integration, results in a series of {\it triangle closure} constraints for the edge vectors $\ell_e$ around the triangles $t\equiv l^*$:
\be
\prod_{t}\delta_{\bb R^4} \bigg(\sum_{e:e \subset t} \epsilon(e|t) \ell'_{e}\bigg) .
\ee
Once again, this procedure does not lead to 10, but only 6 such delta functions. On-shell of 1-flatness, which is imposed in \eqref{eq:discreteZ3} by the first series of delta functions, these six closure are enough to deduce the other four\footnote{However, off-shell of the flatness constraint, it is possible to deduce the 4 missing  triangle closures only in a form, which  involves an unwanted amount of parallel transports.}.

This leads us to write the BFCG-Yetter boundary amplitude formula in the form
\begin{align}
\label{eq:discreteZ4}
Z_Y(\sigma|\tilde\Psi_\Gamma) =
\int \DD h_{l} \DD \ell_{e}  
\prod_{f\in\Gamma} \delta_{\SO(4)}\bigg( \overleftarrow{\prod_{l:l\in\pp f}} h_{l}^{\epsilon(l|f)} \bigg)  
\prod_{t:t\in\pp\sigma} \delta_{\bb R^4} \bigg(\sum_{e:e\in t} \epsilon(e|t) \ell'_{e}\bigg)
\tilde \Psi_\Gamma(h_{l},\ell_{e}).
\end{align}
Henceforth, we take this formula as the definition of the BFCG-Yetter amplitude for a boundary wave function $\tilde\Psi_\Gamma(h_l, \ell_e)$ with arguments the discrete 1-connection $h_l\in\SO(4)$ and edge vectors $\ell_e\in\bb R^4$.
In particular, we have to provide a precise definition for the parallel transport convention of the edge vectors -- a definition that must be compatible with the 1- and 2-gauge symmetry of the problem. This choice is relevant only off-shell of the 1-flatness constraint, where different choices lead to distinct definitions -- not necessarily off-shell equivalent to that obtained by integrating out the bulk variables.

For now, we observe that the triangle closures appearing in the delta-functions correspond to a direct discretization on 2-dimensional objects (triangles) of the 2-Gauss, or torsionfreeness, constraint \eqref{eq:torsionless}. 
However, in a more general setup where the flatness constraint is not implemented, triangle closures will not be enough to fully implement the discrete 2-Gauss constraints: edge simplicity will be required.
In the next section, without entering into the details of choosing a convention for the parallel transports, we will  focus on general considerations that will take us directly to the introduction of the edge simplicity constraints.

\subsection{Boundary edge simplicity\label{sec:boundaryedgesimplicity}}

As we have already noticed, \eqref{eq:discreteZ4} features two classes of boundary variables.
First, there are the four-vectors $\ell_{f}\in\bb R^4$.
They are associated to the  faces of $\Gamma$, that is to the edges in the boundary triangulation of the 4-simplex $\sigma$ (see Table \ref{tab:Gamma}).
A given edge vector is necessarily defined with respect to an (orthonormal) frame. 
As we saw, each tetrahedron in the boundary triangulation is equipped with such a frame. 
Now, changes in such frames when going from one tetrahedron to the neighbouring one is described by the parallel transports $h_l\in\SO(4)$.
Indeed, the group elements $h_l$ are associated to the links in the boundary dual complex $\Gamma$, and hence to triangles in the boundary triangulation. 
They connect two tetrahedral orthonormal frames, and provide the change in this frame when going from one tetrahedron to a neighbouring one across the unique triangle they share. 

Thus, to make the boundary data well defined, we must 
(\textit{i}) choose for each edge $e$ a tetrahedron providing the frame in which the edge vector is expressed.
Reasonably, this tetrahedron should be sharing the given edge. 
Of course, to express the triangle closures, we will need to eventually parallel transport a given edge vector to all three tetrahedra, that share this edge. 
Therefore, we must also 
(\textit{ii}) fix a prescription for the path defining this parallel transport. 
Again, for simplicity, we can demand that this path goes only through the three tetrahedra that share the edge $e$.
Still, we are left with two choices: namely between the clockwise or anticlockwise path around $e$.
The two resulting edge vectors will then differ by the action of the holonomy around the given edge, i.e. 
\be
h_{e^*} \equiv {\overleftarrow\prod}_{l:l\in\pp f \atop \text{for }f =e^*} h_l^{\epsilon(l|f)}.
\ee 

Clearly, the choice of these paths is irrelevant on-shell of the 1-flatness constraints. 
However, in the following sections it will become clear that we need an expression for the triangle closure constraints that is unambiguous also in the presence of curvature.
Indeed, this will be required not only to rewrite the partition function \eqref{eq:discreteZ4} in terms of 2-group representation data and thus prove the relationship between the BFCG-Yetter and the KBF models, but also to construct a boundary Hilbert space for the states $\Psi$'s that supports curvature excitations while still having the 1-Lorentz invariance and triangle closures both implemented.

To solve these problems it is enough to impose a constraint that ensures that the choice for the parallel transport described above does not matter.
This constraint demands that the holonomy $h_{e}$ around an edge $e$ (with starting point in an adjacent tetrahedron to $e$) stabilizes the associated edge vector $\ell_{e}=\ell_{f^*}$ (given in the frame of the same adjacent tetrahedron):
\be
\label{eq:edgesimpli1}
\text{edge simplicity: }\quad h_{e^*} \act \ell_{e}\,=\, \ell_{e}  .
\ee
This condition is known as {\it edge simplicity} \cite{Dittrich:2008ar,Dittrich:2010ey,Dittrich:2012rj}.

Edge simplicity can be interpreted in several ways.
Firstly, it can be interpreted as the minimal part of the flatness constraint necessary to make the edge vectors and the triangle closures well defined independently of a choice of parallel transport \cite{Waelbroeck:1993sm,Dittrich:2008ar}. 
Secondly, and in line with our introduction of these constraints in the continuum \eqref{sec:contedgesimplicity}, edge simplicity can also be understood as the condition the discrete connection $h_l$ has to satisfy in order to be Levi-Civita \cite{Immirzi:1994gq,Dittrich:2010ey}. A connection is Levi-Civita if and only if it is metric and torsion free: in this sense edge simplicity can be understood also as a component of the discrete 2-Gauss constraint (torsionfreeness, see discussion before \eqref{eq:contedgesimplicity}).

In a 4d triangulation, the discrete Levi-Civita requirement is that the holonomy around a triangle leaves the plane of this triangle point-wise invariant. 
Here we work with  the boundary triangulation which is three-dimensional, and therefore we consider holonomies that leave edges invariant, rather than planes. 
However, notice that these edge vectors are $\bb R^4$ vectors and that an $\SO(4)$ rotation that leaves a 4-vector invariant necessarily leaves a whole plane (on which this 4-vector lies) point-wise invariant too\footnote{The reason is that the eigenvalues for an $\SO(4)$ rotation come in complex conjugated pairs. More geometrically, think of a $\SO(4)$-rotation that leaves the ``time'' vector (1,0,0,0) invariant: it reduces to an $\SO(3)$ rotation of the ``spacelike'' hyperplane, and any such 3d rotation has a fixed vector.}.
{ We can interpret this plane to be the one given by the 4d triangle obtained by linking the boundary edge $e$ to the center of the 4-simplex $\sigma$.}
It is thus sufficient to impose equation \eqref{eq:edgesimpli1} on the $\SO(4)$ holonomies in order to ensure the 4d Levi-Civita condition.

To conclude, let us mention that the name ``edge simplicity" arose in the context of the Hamiltonian analysis of the ``simplicity constraints'' that reduce the phase space of 4d BF theory to that of Palatini-Plebanski gravity, by demanding that the BF bivector $B$ 2-form is of the so-called ``simple'' kind, that is $B=\star E\wedge E$ for some vector 1-form $E$ (see discussion in  the conclusion section \ref{outlook}).
More specifically, one can  show that the edge simplicity constraints represent a discretization of the secondary simplicity constraints that arise for the connection variables \cite{Dittrich:2010ey}.
Consistently, in the next section we will show that the simplicity constraint \eqref{eq:edgesimpli1} are indeed partially fixing the holonomy variables.

\subsection{From edge simplicity to discrete Levi-Civita holonomies}\label{sec:getu1}

Here we are going to show that the edge simplicity constraints restrict the holonomy variables $h_{e^*}$ to one free parameter per link\footnote{The discussion of this section can be easily generalized to arbitrary boundary triangulations $\Delta_3$, more general than $\Delta_3=\pp\sigma$.}.  
To this end, denoting by $\tau$ the tetrahedra in $\pp\sigma$, we assume that:
(\textit{i}) the $\ell_e[\tau]$'s are given to us;
(\textit{ii}) they are related to each other by $\ell_e[\tau'] = h_{l=t^*}\triangleright \ell_e[\tau]$ where $t=\pp \tau_1 \cap \pp \tau_2$ is the triangle shared by two neighbouring tetrahedra sharing the edge $e$, i.e. $e\in\pp t$;
(\textit{iii}) the $\ell_e[\tau]$'s satisfy the triangle closure conditions at each $\tau$ and therefore define a geometric tetrahedron in that frame;  
(\textit{iv}) the five tetrahedra so defined are nondegenerate and therefore define (up to a sign) a unit normal vector $n_\tau[\tau]\in\bb R^4$. { This vector is uniquely defined once a topological orientation for $\pp\sigma$ is chosen. We assume this choice has been made and $n_\tau[\tau]$ is positively oriented. Hereafter, we will consistently denote in square brackets the reference frame in which a quantity is defined.}

The first, crucial, step is to notice that condition (\textit{ii}) is consistent precisely thanks to the edge simplicity constraints. 
As a second step, we decompose $h_{t^*}$ into a part which is fully fixed by its parallel transport properties, and a part that contains the remaining freedom present in $h_{t^*}$.
For his purpose, consider a triangle $t$, and in it a pair of edges $e_1$ and $e_2$ (which pair of edges is chosen will not matter, thanks to the triangle closure constraint). Also, denote by $\tau'$ and $\tau$ the two tetrahedra  that share $t$.
The edge vectors of $e_1$ and $e_2$ must then be related by
\be
\ell_{e_1} [\tau'] = h_{t^*} \act \ell_{e_1}[\tau]
\quad\text{and}\quad
\ell_{e_2} [\tau'] = h_{t^*} \act \ell_{e_2}[\tau].
\label{eq:Edge60}
\ee
This fact, together with (\textit{iv}) above, allows us to decompose $h_t$ into two $\SO(4)$ elements:
\be
h_{t^*} = {\cal B}_{t^*} {\cal R}_{t^*},
\label{eq:hdecompos}
\ee
where ${\cal R}_{t*}={\cal R}_{t^*}(\ell[\tau],\ell[\tau'], n_\tau, n_{\tau'})$ is uniquely defined by the condition that it must properly transport the triangle $t$ as well as the tetrahedron normal\footnote{That this rotation exists and is unique, can be most easily seen by working in ``time gauge'' i.e. by appending to it two $\SO(4)$ rotations $g_\tau$ and $g_{\tau'}$ that take $n_\tau[\tau]$ and $n_{\tau'}[\tau']$ onto $(1,0,0,0)$. Then one reduces the problem to a 3d problem, whose solution is evident.} $n$:
\be
\ell_{e_1} [\tau'] = {\cal R}_{t^*} \act \ell_{e_1}[\tau],
\quad
\ell_{e_2} [\tau'] = {\cal R}_{t^*} \act \ell_{e_2}[\tau],
\quad\text{and}\quad
n_{\tau'} [\tau'] = {\cal R}_{t^*} \act n_{\tau}[\tau].
\label{eq:Edge61}
\ee

The third and last step consists in constraining the form of the ``boost'' part ${\cal B}_{t^*}$ of $h_{t^*}$. 
As $h_{t^*}$ satisfies \eqref{eq:Edge60} and ${\cal R}_{t^*}$ satisfies \eqref{eq:Edge61}, ${\cal B}_{t^*} = h_{t^*} {\cal R}_{t^*}^{-1}$ must pointwise stabilize the triangle $t$ in the frame of $\tau'$:
\be
\ell_{e_1}[\tau']={\cal B}_{t^*} \act \ell_{e_2}[\tau'] 
\q\text{and}\q
\ell_{e_2}[\tau']={\cal B}_{t^*} \act \ell_{e_2}[\tau'].
\ee
Now, an $\SO(4)$ rotation that stabilizes two linearly independent four-vectors is necessarily a simple rotation that rotates the plane normal to that $t$ by a free angle $\theta_t$. 
Using the explicit value of the generators $J_{ab}$ as given in  \eqref{eq:so4generator}, we have 
\be
{\cal B}_{t^*} = \exp \left( \theta_t \;\frac{\star (\ell_{e_1}[\tau'] \wedge \ell_{e_2}[\tau'])\cdot J}{2 A_t }  \right),
\label{eq:boost}
\ee 
where $\star$ is a dualization of the internal indices, i.e. $\star( \ell\wedge \ell')^{ab} := \tfrac12 \epsilon^{abcd}\ell_{[c}\,\ell'{}_{d]} $, and thus $\star (\ell\wedge \ell')\cdot J = \tfrac14 \epsilon^{abcd}\ell_{[c}\,\ell'{}_{d]} J_{ab}$.

Hence, we conclude that on-shell of the edge simplicity constraint, $b_{t^*}$---and therefore $h_{t^*}$---has a single free parameter $\theta_t$ corresponding to the ${\rm U}(1)$ stabilizer group of the edge vectors spanning the triangle $t$.

\medskip

To understand this angle from a geometrical perspective, it is enough to notice that $n_\tau[\tau]$ lies in the plane orthogonal to $\ell_{e_1}[\tau]$ and $\ell_{e_2}[\tau]$.
Hence, from the above discussion, we see that ${\cal B}_{t^*}$ rotates $n_{\tau'}[\tau']$ by the angle $\theta_t$ in the plane orthogonal to  $\ell_{e_1}[\tau]$ and $\ell_{e_2}[\tau]$.
This means, in turn, that $h_{t^*}$ rotates the tetrahedron 4-normal $n_\tau[\tau]$ by an angle $\theta_{t^*}$ {\it with respect to} $n_{\tau'}[\tau']$ in the same plane, since
\begin{align}
n_\tau[\tau'] := h_t \act n_\tau[\tau] &= ({\cal B}_{t^*} {\cal R}_{t^*}) \act n_\tau[\tau] = {\cal B}_{t^*} \act n_{\tau'}[\tau'] 
\end{align}

Now, since the (cosine of the) exterior dihedral angle between two tetrahedra in the boundary of a four-simplex is given by the inner product between the two normals, expressed in the {\it same} reference frame, the above discussion leads us to 
\be\label{thetadef0}
n_{\tau'}[\tau'] \cdot  n_\tau[\tau'] = n_{\tau'}[\tau'] \cdot ( {\cal B}_{t^*} \act n_{\tau'}[\tau'] )  = \cos\theta_t,
\ee
from which we conclude that: 
{\it after the complete imposition of the discrete 2-Gauss constraint in the form of triangle closures {\it and} edge simplicity constraints, the holonomies $h_{l=t^*}$ are said to be Levi-Civita and they feature only one free parameter each; this parameter, denoted $\theta_t$, has the geometric interpretation of the exterior dihedral angle between the tetrahedra $\tau$ and $\tau'$ hinging at the triangle $t$.}

\subsection{1-flatness for Levi-Civita holonomies \label{sec:1flatness4LC}}

In the previous sections, we showed that full implementation on $\pp \sigma$ of the discrete 2-Gauss constraint -- as encoded in both triangle closures and edge simplicity -- leads us to consider 5 geometric tetrahedra defined by the edge vectors $\ell_e[\tau]$ whose reference frames are related by holonomies
\be
h_{l=t^*} =  \exp \left( \theta_t \;\frac{\star(\ell_{e_1}[\tau'] \wedge \ell_{e_2}[\tau'])\cdot J}{2 A_t }  \right) {\cal R}_{t^*},
\ee
with only one free parameter, the 4d exterior dihedral angle $\theta_t$ (recall that ${\cal R}_{t^*}$ is fully determined by the set of $\ell_e[\tau]$, cf. equations \eqref{eq:hdecompos}, \eqref{eq:Edge61}, and \eqref{eq:boost}).

Thus, we can interpret $\pp \sigma$ as a discrete three-dimensional hypersurface whose intrinsic geometry (induced metric) is fully determined by the assignment of the edge vectors $\ell_e[\tau]$ and whose extrinsic geometry (extrinsic curvature) is encoded in the exterior dihedral angle $\theta_t$. 

Clearly, if we knew that the embedding space for $\pp\sigma$ was flat, a discrete analogue of the Gauss-Codazzi equation would tie the extrinsic and intrinsic geometry together. In other words, knowing that $\pp\sigma$ is embedded in flat 4-space, allows us to compute all its intrinsic and extrinsic properties from its edge-lengths only: e.g. knowledge of a flat embedding allows us to determine the {\it exterior} 4d dihedral angles $\theta_t=\Theta_t(\{\ell_e[\tau]\})$ in terms of the 3d {\it interior} dihedral angles $\phi_e[\tau]=\phi_e[\tau](\{\ell_{e'}[\tau]\})$ intrinsic to $\pp\sigma$, and vice versa, according to the well-known formulas\footnote{These formulas follow from the spherical law of cosines obtained by drawing a unit sphere around one of the vertices of the simplex under consideration. For a proof and discussion of further implications we refer the reader  to e.g.  \cite{Dittrich:2007wm,Dittrich:2008va,Baratin:2014era}.}
\begin{align}
\label{eq:flatness63}
\begin{dcases}
\cos \Theta_{kl} \,=\, - \epsilon \frac{\cos \phi_{kl}[i] - \cos \phi_{ki}[l] \cos \phi_{li}[k]}{\sin \phi_{ki}[l] \sin \phi_{li}[k]}  ,\\
\cos \phi_{kl}[i] \,=\, -\epsilon  \frac{\cos \Theta_{kl} - \cos \Theta_{ik} \cos \Theta_{il}}{\sin \Theta_{ik} \sin \Theta_{il}}  ,
\end{dcases}
\end{align}
where we have labelled the tetrahedra of the 4-simplex by $i,j,\dots\in\{1,2,\dots,5\}$ and denoted: by $\Theta_{kl}$ the exterior 4d dihedral angle between the tetrahedra $k$ and $l$, and by $\phi_{kl}[i]$ the internal 3d dihedral angle in $i$ between the triangles that $i$ shares with $k$ and $l$.   
Also, we have introduced a sign $\epsilon=\pm1$ corresponding to a {\it global} choice of orientation that is not fixed by the relations between the angles.
 
This is precisely the role of the 1-flatness constraint: it tells us that 4d space is flat, and that -- on-shell of it -- the 4d dihedral angles $\theta_t$ are hence fully determined to be the geometric dihedral angles $\Theta_t(\{\ell_e[\tau]\})$ as in \eqref{eq:flatness63}.

\section{Boundary states: from spin-network to G-network functions}\label{sec:5}

As recalled in section \ref{sec:BF2PR}, given a boundary state  $\Psi_\Gamma$ of 3d BF theory, its associated amplitude is given by 
 \be
Z_{BF}(\tau|\Psi_\Gamma) = \int \DD h_{l}  \, Z_{BF}(\tau|h_{l}) \Psi_\Gamma(h_{l}).
\label{eq:Z_BFPsi-bis}.
\ee 
There we also recalled that any $\Psi_\Gamma$ that satisfies the Gauss constraint, i.e. any $\Psi_\Gamma$ that is Lorentz invariant,\footnote{i.e. $\Psi_\Gamma( g_{t(l)}^{-1} h_l g_{s(l)}) = \Psi_\Gamma(h_l)$ with $t(l)$ and $s(l)$ being respectively the target and the source nodes of the link $l$.} can be decomposed via Peter-Weyl theorem onto the spin-network basis.
This basis is labelled by irreducibile representations of $\SU(2)$ (spin $j_l\in\frac12\bb N$) associated to its links, and by $\SU(2)$-invariant tensors (intertwiners) implementing the Gauss constraint associated to its nodes \cite{Rovelli:1995ac}. 
Using this decomposition, the amplitude (\ref{eq:Z_BFPsi-bis}) is readily mapped onto the Ponzano-Regge amplitude \eqref{eq:BFj}:
\be
Z_{BF}(\Psi_\Gamma) =   \sum_{\{j_e\}} \mu_{j_e}\{ 6 j_e \} \tilde\Psi_\Gamma(j_e).
\label{eq:BFj-bis}
\ee
Notice that geometrically the spins represent the lengths of the boundary edges dual to the spin-network's links, and that the 3-valent intertwiners associated to the tetrahedral spin-network are unique. In this sense, the spin-network basis can be interpreted as a basis of 2d discrete quantum geometries.

We  would like to  generalize this chain of results to the four-dimensional case: mapping BFCG-Yetter onto KBF amplitudes.
The BFCG-Yetter amplitude of a boundary state $\tilde\Psi_\Gamma$ is given in this case by
\begin{align}
\label{eq:discreteZ4p}
Z_Y(\sigma|\tilde\Psi_\Gamma) =
\int \DD h_{l} \DD \ell_{f}  
\prod_{f\in\Gamma} \delta_{\SO(4)}\bigg( \overleftarrow{\prod_{l:l\in\pp f}} h_{l}^{\epsilon(l|f)} \bigg)  
\prod_{t:t\in\pp\sigma} \delta_{\bb R^4} \bigg(\sum_{e:e\in t} \epsilon(e|t) \ell'_{e}\bigg)
\tilde \Psi_\Gamma(h_{l},\ell_{e}).
\end{align}
It is natural to conjecture that using a generalization of the Peter-Weyl theorem for 2-groups to decompose $\tilde \Psi_\Gamma$ onto 2-representations and imposing the 1- and 2-Gauss constraints would give us the sought relation between the BFCG-Yetter and the KBF amplitudes. 
However, to our knowledge there is no analogue of the Peter-Weyl theorem for 2-groups.
For this reason, we will follow a different route: we will first impose discrete versions of the 1- and 2-Gauss constraints studied in the previous sections on $\tilde \Psi_\Gamma$ and hence follow the quantum-geometric intuition they provide to rewrite the BFCG-Yetter amplitude in a way that manifestly matches the KBF amplitude. 
This geometric route is available also in 3d BF theory, and will be briefly reviewed in the next subsection.

In the case of the BFCG-Yetter theory, we will find that all boundary states that satisfy our proposed discrete version of the 1- and 2-Gauss constraints -- i.e. 1-Lorentz invariance at each tetrahedron together with triangle closures and edge simplicity -- can be expanded on a basis of states labeled by edge lengths $l_e:=|\ell_e|\in\bb R_+$ and integers $s_f\in\bb Z$. These labels match the labels proper of the theory of irreducible 2-representations of the Poincar\'e 2-group \cite{Baez:2008hz,Baratin:2014era}. However, our result will be derived from purely geometrical considerations, and not algebraically from an (unknown) generalization of the Peter-Weyl theorem to functions of 2-group elements. 
Hence, we refer to our proposed basis of boundary states as {\it G-networks}\footnote{As opposed to 2-spin-networks, a name that we feel bring back an association to the Peter-Weyl theorem.} -- the letter G standing for ``Gauss''.

We will also show that, dually, G-networks can be understood as a basis of the space of functions $\Psi_\Gamma(h_\ell, x_f)$ that are 1- and 2-Lorentz invariant, cf. \eqref{eq:Ftrafo0}.

\subsection{Spin networks for 3d BF theory from discrete 3d geometries}\label{3ddiscrete}

Before discussing the BFCG case, let us recall how in 3d BF theory one can characterize the boundary states from the discrete geometry picture.

Recall that here $\Gamma$ is the dual of a discretization of the 2d boundary. The details of the discretization have been discussed most recently in \cite{Dupuis:2017otn}.

Upon discretization, one associates to each edge of $\Gamma$ the phase space $\mathrm T^*\SU(2)\cong \SU(2)\times \su^*(2)\ni (h_l,y_l)$. 
In particular, the holonomy $h_l$ is associated to the (oriented) links $l\in\Gamma$, and the flux $y_l\equiv y_e^{(s)}$ is associated to the source node of $l=e^*$, i.e. $n=s(l)$.
To the target node, $n=t(l)$ one associates the parallel transported flux $ y_e^{(t)}\equiv - h_l y_e h_l^{-1}$. 
The fluxes $y_e^{(s)}, \, y_e^{(t)}\in \su^*(2)\cong \bb R^3$ geometrically represent the (oriented) 3d edge vector associated to the edge $e=l^*$ (recall that we are in 2d, see table \ref{tab:Delta2}).\footnote{Here we have implicitly identified $\su^*(2)$ with $\su(2)$ via the Killing form. Using a matricial representation we have represented the natural (co)adjoint action in terms of matrix conjugation. Identifying $\su^*(2)\cong \bb R^3$, the coadjoint action translates into a 3d rotation.}

In this discrete geometry picture, the discretization of the constraints is straightforward. The Gauss (or torsionfreeness) constraint is imposed at each node $n\in\Gamma$, whereas the flatness constraint is imposed at each face $f\in\Gamma$,
\be
T=0 \leadsto  \cG_n= \sum_{l: \, n\in\partial l} y_{l}^{\epsilon(n|l)}=0, 
\quad
F=0\leadsto \cF_f = \prod_{l_i\in \partial f} h_{l_i} - \bb 1_{\SU(2)} = 0,  
\label{eq:3dconstraintsGF}
\ee
where $\epsilon(n|l)$ is either $s(l)$ or $t(l)$ and keeps track of the orientation of the link. 

Notice that $\mathrm T^*\SU(2)$ is equipped with a canonical symplectic structure.
With respect to this symplectic structure, the discretized Gauss and flatness constraints are first class and reflect their continuum algebra. 
The Gauss and flatness constraints generate Lorentz transformations at the nodes and translations of the vertices respectively. 
In this sense, it is natural to see the flatness constraints as implementing the diffeomorphism symmetry, see \eqref{eq:3dBFdiffeo} and the related discussion.

As stated above, we want to study the space of states that satisfy the Gauss constraint, but not necessarily the flatness constraint.\footnote{These states can be expanded on the spin-network basis, and are said to be ``kinematical'' -- this nomenclature is inherited from 4d quantum gravity, where the flatness constrained is replaced by a more involved Hamiltonian constraint, while the Gauss constraint is left untouched. For a discussion of the dual choice in 3d gravity, see \cite{Dupuis:2017otn}.}
Rather than following the usual Peter-Weyl route, let us consider the geometrical picture.

The 2-complex $\Gamma$ is dual to a 2d (boundary) triangulation, i.e. $\Gamma = \Delta_2^*=(\pp \Delta_3)^*$.
The edge vectors of $\Delta_2$ are given by the $y_e$.
For the edge vectors to form triangles, we need to impose a constraint asking their (oriented) sum at each node to vanish. This is nothing else than the Gauss constraint ${\cal G}_n$ \eqref{eq:3dconstraintsGF}.
Moreover, each node/triangle has its own reference frame, and it is the role of the holonomy $h_l$ to perform the parallel transport between such reference frames, and specifically from the node $n=s(l)$ to $n'=t(l)$. 
The action of these parallel transports allows to fully reconstruct the triangulation $\Delta_2$. 
Notice that the flatness constraints then encode the vertex-translation invariance (diffeomorphism symmetry)  and specify the dynamics. In particular it can be found to fix the extrinsic geometry of $\Delta_2=\pp\Delta_3$ in $\Delta_3$ understood as a (discrete) Euclidean flat space. 

Let us formalize this construction.
Choosing the connection polarization for the boundary state, we obtain the Hilbert space
 \be
{\cal L}^2\Big( \bigotimes_{l} \SU(2) \Big)\ni \Psi_\Gamma(h_{l}).
\label{eq:64hilbert}
 \ee   
 On this Hilbert space, the matrix elements  $(h_{l_o})^A{}_B$ of the $l_o$-holonomies act as multiplication operators. 
 The fluxes $y_{e_o}^{(s)}$ (resp. $y_{e_o}^{(t)}$) act instead as left-invariant Lie derivatives $i (L_{l_o})^{a}$ on the $(l_o=e^*_o)$-th argument (resps. as right-invariant Lie derivative):
\be
(L_{l_o})^{a}\, \Psi_\Gamma(h_{l}) \,:=\, \frac{\d}{\d t}_{|t=0}  \Psi_\Gamma(h_{l\neq l_o}, h_{l_o} e^{ t J^{a}}),  
\label{eq:65}
\ee
where  $\{J^a\}_{a=1,2,3}$ are the generators of $\su(2)$.  
Then, the Gauss constraint acts as (for notational simplicity we assume all the links at the node $n$ to be outgoing):\footnote{For completeness, we report also the action of the flatness constraint, $$\left({\cal F}_f \right)_{AB} \, \Psi_\Gamma (h_l) \,=\, \left(\Big( {\prod}_{l: l\in \partial f} h_{l}\Big)_{AB} - \delta_{AB}\right) \Psi_\Gamma  (h_l).$$}
\begin{align}
\left({\cal G}_n\right)^a  \, \Psi_\Gamma (h_l) &\,=\, i \left( {\sum}_{l: \, n\in\partial l}  L_{l}^{a} \right) \Psi_\Gamma  (h_l)
\end{align}
Since from \eqref{eq:65} it follows that the Gauss constraints implements Lorentz invariance at the nodes, we see that implementing it means that we have to retain only the Lorentz invariant information encoded in the edge data $y_e\in\su(2)\cong\bb R^3$: this is the edge lengths $l_e = \sqrt{y_e\cdot y_e}$. 
Since $l_e^2$ is the quadratic Casimir of $\su(2)$, we see that the spectrum of the $l_e$ -- labelling a basis of the $\Psi_\Gamma$ satisfying the Gauss constraints -- is indexed by the irreducible representations of $\SU(2)$, $j_e\in\frac12\bb N$. 
Notice that a triangulation where only the edge-lengths are specified has the same input data as in Regge calculus, of which the Ponzano-Regge model constitutes a quantization (in 3d). 
For a detailed discussion on the reduction of the dynamics of 3d LQG to Regge calculus, see \cite{Bonzom:2013tna}.    

Let us now extend this construction to the 4d case.

\subsection{G-networks for the BFCG-Yetter model from discrete 4d geometries }

In this section we will develop a discretized phase space for the boundary data of the BFCG-Yetter model with the Poincar\'e gauge 2-group.
In particular we will translate the geometric constructions of the previous sections -- that is triangle closures, edge simplicity, 1-flatness, etc. -- into the language of constraints on the phase space and their quantization in a Schr\"odinger representation.
We will keep focusing on the case of a single 4-simplex, which we will analyze in great detail.

Of particular interest will be the edge-simplicity constraint, which is key to relate the BFCG-Yetter theory to the KBF state sum, but also the 2-flatness constraints,  which we will interpret as a combined diffeomorphism and Hamiltonian constraint at the light of the fact that its action generates 4d translations of the vertices in the triangulation.

\subsubsection{The boundary of a 4-simplex: conventions and notation}

The boundary of the 4-simplex $\sigma$ is a discretization of the three-sphere $S_3$ into 5 tetrahedra, each pair of which shares one triangular face. 
We label the 5 vertices $v$ of the four simplex by $\bar i, \bar j, \dots \in \left\{ \bar 1, \bar 2, \dots, \bar 5\right\}$, so that its 10 edges are labeled by pairs $e\in\left\{\left[\,\overline{i j}\,\right]\right\}$, its 10 triangles by triples $t\in\left\{\left[\, \overline{ i  j k}\,\right]\right\}$, and its 5 tetrahedra by quadruplets $\tau\in\left\{\left[\,\overline{ i j k l}\,\right]\right\}$. 

\begin{figure}[t!]
\begin{center}
\begin{tikzpicture}[scale=0.65]%
\def\ra{4.2}
\def\rb{2.7}
\begin{scope}
\foreach \x in {18,90,...,306} {
   	\draw (\x:\ra) node{} -- (\x+72:\ra) ;
       	\foreach \y in {18,90,...,306}{
		\draw (\x:\ra) -- (\y:\ra) ;
		} ;
};
\node[right] at (18:\ra) {$\bar{1}$};
\node[above] at (90:\ra) {$\bar{2}$};
\node[left] at (90+72:\ra) {$\bar{3}$};
\node[below left] at (90+2*72:\ra) {$\bar{4}$};
\node[below right] at (90+3*72:\ra) {$\bar{5}$};
\end{scope}
\begin{scope}[rotate = 35,red!60]
\foreach \x in {18,90,...,306} {
   	\draw (\x:\rb) node{} -- (\x+72:\rb) ;
       	\foreach \y in {18,90,...,306}{
		\draw (\x:\rb) -- (\y:\rb) ;
		} ;
};
\node[right] at (18:\rb) {4};
\node[above] at (90:\rb) {5};
\node[left] at (90+72:\rb) {1};
\node[below left] at (93+2*75:\rb) {2};
\node[below right] at (93+3*75:\rb) {3};
\end{scope}
\end{tikzpicture}
\end{center}
\caption{$\{\b1,\b2,\b3,\b4,\b5\}$ are the vertices of the 4-simplex $\sigma$, joined by black lines corresponding to the edge of the 4-simplex. 
In red, we represented the dual {\it boundary} graph $\Gamma=(\pp\sigma)^*$.}\label{penta}
\end{figure}
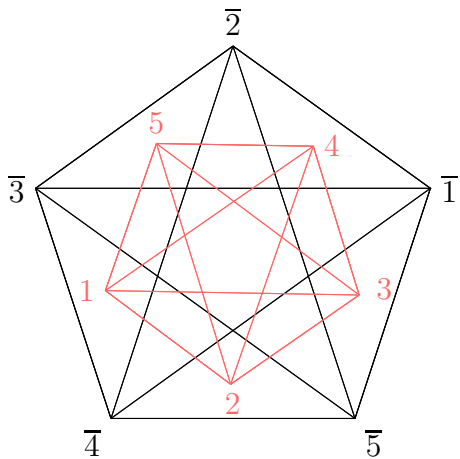

We similarly label the Poincar\'e dual of the 4-simplex.
Its 5 dual nodes $n$ are labeled by $i,j,\dots \in \{ 1,2,\dots,5\}$ and correspond to the boundary tetrahedra which do not contain the barred version of the $i$ label (e.g. $1 \leftrightarrow [\overline{2345}]$); similarly, the 10 dual links corresponding to the boundary triangles are $l\in\{[ij]\}$, the 10 dual faces corresponding to the boundary edges are $f\in\{[ijk]\}$, and finally the 5 dual ``bubbles'' (3-cells) corresponding to the boundary vertices of the four simplex are $b\in\{[ijkl]\}$. See figure \ref{penta} and table \ref{tab:Gamma}.

For convenience, when there is no risk of confusion, we will sometimes label cells in the 4-simplex (resp. of its dual) with the dual label (resp. with the direct label).  

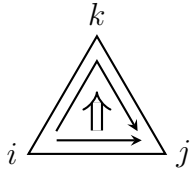
\begin{figure}[ht!] 
\centering
\begin{tikzpicture}[scale= 0.6 , arrow1/.style={thick,->,shorten >=2pt,shorten <=2pt,>=stealth},
      arrow2/.style={thick,double distance = 3pt,-implies,shorten >=2pt,shorten <=2pt}  ] 

\coordinate (A) at (0,0);
\coordinate (B) at (3,0);
\coordinate (C) at (1.5,2.6);

\draw[thick] (A)--(B)-- (C) --cycle ;
\draw[arrow1] (0.55,0.4)--(1.5,2.05)--(2.45,0.4); 
\draw[arrow1]  (0.5,0.3) --(2.6,0.3) ;
\draw[arrow2](1.5,0.4)--(1.5,1.5) ;

\node[left] at (A) {$i$};
\node[right] at (B) {$j$};
\node[above] at (C) {$k$};

\end{tikzpicture}
\caption{The thin and double-lined arrows graphically represent the definition of the 1- and 2-holonomy $h_{ji}$ and  $x_{ijk}$ respectively.}\label{para}
\end{figure}

Next we introduce (boundary) variables $h_{l}$ and $x_{f}$.  For each dual  link $l=[ji]$ with $i<j$  we associate a group element $h_{ji}\in \SO(4)$ which represents the parallel transports from $i$ to $j$ (composition is from the right to the left). We furthermore define $h_{ij}:=h_{ji}^{-1}$.  
Likewise, for each dual face $f=[kji]$, with $i<j<k$ we introduce $x_{kji}\in \mathbb{R}^4$.  This variable $x_{kji}\in \mathbb{R}^4$ represents the 2-parallel transport from the link $[ji]$ to $[jk] \circ [ki]$. See figure \ref{para}. 

We furthermore specify the node $i$ as the root for $x_{kji}$, that is we take $x_{kji}$ to be defined in the reference system of the smallest label $i$. 
For the 2-holonomies $x_{kji}$ there are two kind of inverses, a horizontal and a vertical one.
The vertical inverse $\ominus _V$ describes the 2-parallel transport from $[jk] \circ [ki]$ to $[ji]$, with root $i$. 
The horizontal inverse $\ominus _H$ describes the 2-parallel transport from $[ij]$ to $[ik]\circ [kj]$, with root $j$.  We have (see figures \ref{inverse1} and \ref{inverse2}):
\be
\ominus _V x_{kji}\equiv -x_{kji} 
\qquad\text{and}\qquad
\ominus _H x_{kji}\equiv- (h_{ji}\act x_{kji})  .
\ee

\begin{figure}[t!] 
\begin{minipage}{.5\textwidth}
\centering

\begin{tikzpicture}[scale= 0.6 ,arrow1/.style={thick,->,shorten >=2pt,shorten <=2pt,>=stealth},
      arrow2/.style={thick,double distance = 3pt,-implies,shorten >=2pt,shorten <=2pt}  ] 

\draw[thick] (0,0)--(3,0)-- (1.5,2.6) --cycle ;
\draw[arrow1] (0.55,0.4)--(1.5,2.05)--(2.45,0.4); 
\draw[arrow1]  (0.5,0.3) --(2.6,0.3) ;
\draw[arrow2] (1.5,1.5)--(1.5,0.4) ;

\node[left] at (0,0) {$i$};
\node[right] at (3,0) {$j$};
\node[above] at (1.5,2.6) {$k$};

\end{tikzpicture}
\caption{Vertical inverse.}\vspace{2.3em}\label{inverse1}

\end{minipage}
\begin{minipage}{.5\textwidth}
\centering
\begin{tikzpicture}[scale= 0.6 ,arrow1/.style={thick,->,shorten >=2pt,shorten <=2pt,>=stealth},
      arrow2/.style={thick,double distance = 3pt,-implies,shorten >=2pt,shorten <=2pt}  ] 

\draw[thick] (0,0)--(3,0)-- (1.5,2.6) --cycle ;
\draw[arrow1] (2.45,0.4)--(1.5,2.05)--(0.55,0.4); 
\draw[arrow1] (2.6,0.3)--(0.5,0.3) ;
\draw[arrow2](1.5,0.4)--(1.5,1.5) ;

\node[left] at (0,0) {$i$};
\node[right] at (3,0) {$j$};
\node[above] at (1.5,2.6) {$k$};

\end{tikzpicture}
\caption{Horizontal inverse. Notice that the root has changed from $i$ to $j$, hence the need for a 1-parallel transport.} \label{inverse2}

\end{minipage}

\end{figure}

For the definition of the 2-flatness constraint, we will also need a prescription of how to change the reference system to one of the other nodes in the dual face $[kji]$ (``whiskering''). Here we use the action  of $\SO(4)$ on $\mathbb{R}^4$, and with $x_{kji}\equiv x_{kji}[i]$ we define   
\be
x_{kji}[j]=h_{ji} \act x_{kji}[i] 
\qquad\text{and}\qquad
x_{kji}[k]=h_{ki} \act x_{kji}[i].
\ee

\subsubsection{The discretized BFCG phase space}\label{discphase}

As discussed in section \ref{sec:diffeos}, from the BFCG action \eqref{eq:BFCG13} we can read off that the pull-back of the 1-connection $a$ is canonically conjugate to the pull-back of the bivector field $b$, and that the pull-back of the 2-connection $\sigma$ is canonically conjugate to the pull-back of the tetrad field $c\equiv e$. 
That is, before imposition of the four constraints, the phase space factorizes into the two phase spaces of the canonical pairs $(a,b)$ and  $(\sigma,c)$. Here we postulate that this split survives the discretization and check the consistency of our assumption. A more detailed derivation of the discrete phase space structures is left to future work.

Analogously to \eqref{eq:AXdiscrete}, the 1- and 2-connection variables $a$ and $\sigma$ give rise upon discretization to the 1- and 2-holonomies $h_{ji}$ and $x_{kji}$ respectively:\footnote{Recall that the prime stands for some appropriate 1-parallel transport convention.}
\be
h_{ji} = \mathrm{Pexp}\int_{[ji]} a \in \SO(4)
\quad\text{and}\quad
x_{kji} = \mathrm{Sexp}\int_{[kji]} \sigma' \in \bb R^4.
\ee 
Conversely, their conjugate variables, the bivector field $b$ and the tetrad $c\equiv e$ respectively, are discretized into
\be
b_{ji} = \int_{[ji]^*}  b' \in \so(4)
\qquad\text{and}\qquad
\ell_{kji} = \int_{[kji]^*} c'\in \bb R^4
\ee
where $[ji]^*=\epsilon_{\bar{m}\bar{l}\bar{k}ji}\left[ \,\bar{mlk} \,\right]$ is a triangle and $[kji]^* = \epsilon_{\bar{m}\bar{l}kji}\left[ \,\bar{ml} \,\right]$ is an edge (without any summation on repeated indices).\footnote{The need for dualization reflects the presence of a {\it wedge} product in the action between $B$ and $\d A$, and between $C$ and $\d \Sigma$.}
Notice that this is compatible with
\be
\ell_{\bar{lm}} = - \ell_{\bar{ml}}
\qquad\text{and}\qquad
b_{ij} = - h_{ji} b_{ji} h_{ji}^{-1},
\ee
where the inverse of $\ell_{\bar{ml}}$ considered here corresponds to the vertical inverse of $x_{kji}$, with no change of the reference frame which for $\ell_{kji}$ is taken to be $i$ (i.e. the same as for its conjugate variable $x_{kji}$). Moreover, $\ell_{kji}$ follows the same parallel transport conventions with the $\SO(4)$ 1-holonomies  as $x_{kji}$.

Thus, the discrete phase space associated to the boundary of the 4-simplex $\Gamma = (\pp \sigma)^*$ is given by a product of link phase spaces $\mathrm T^*\SO(4) \cong \SO(4)\times \so(4) \ni (h_{ji},b_{ji})$ and of face phase spaces $\mathrm T^*\bb R^4 \cong \bb R^4\times \bb R^4 \ni (x_{kji}, \ell_{kji})$. 
The non-vanishing Poisson brackets of this phase space are
\be
\{ b_{ji}^{A} , (h_{ji})^c{}_d \} = (h_{ji} J^{A})^c{}_d
\qquad\text{and}\qquad
\{ b_{ji}^{A} , (b_{ji})^{B} \} = f^{AB}{}_{C} b^{C}_{ji},
\ee
and 
\be
\{ \ell^a_{kji} , x_{kji}^b \} = \delta^{ab}.
\ee
where we introduced the superindex $A=[ab]$ to label a basis of $\so(4)$; thus, $f^{AB}{}_{C}$ are the structure constants of $\so(4)$, as in $[ J_{A}, J_{B}] = f_{AB}{}^C J_C$.
Notice that the discretization of the $a$- and $b$-fields, and their phase space structure, reproduces the discretization of BF theory described in section \ref{3ddiscrete}.

\subsubsection{The discretized constraints and their action}\label{sec:constraints}

In this section we present the discrete constraints for 1- and 2-Lorentz and shift symmetry, as well as the symmetry transformations that they generate on the 4-simplex data through the Poisson algebra introduced in the previous section.
The analogy with the continuous structures of section \ref{sec:diffeos} will be evident, even when it will be fogged by the necessity to parallel transport all the relevant quantities from their roots to the point where the action takes place. 

In table \ref{tab:cons-where}, we provide a summary of where the different constraints act.
\begin{table}[t]
\begin{center}
\begin{tabular}{c||c}
\hline
 \bf Constraint    & \bf Action locus                  \\ \hline
1-Gauss  & node/tetrahedron 
\\ \hline
2-Gauss & links/triangle 
  \\ \hline
1-flatness & face/edge 
\\ \hline
2-flatness  & bubble/vertex
\\ \hline
\end{tabular}
\caption{Action loci of the discrete constraints.}\label{tab:cons-where}
\end{center}
\end{table}

\paragraph*{1-Gauss for 1-Lorentz}
The 1-Gauss constraints \eqref{1-GaussC} generates 1-Lorentz transformations \eqref{eq:12Lorentz}.
In the continuum these are parametrized by $\SO(4)$-valued 0-forms. 
Hence, upon discretization, the Gauss constraints becomes an $\so^*(4)$-valued quantity based at the nodes of $\Gamma$:\footnote{This expression comes from the following one, once $(J_{ab})_{cd}$ is replaced by its explicit matricial form \eqref{eq:so4generator},  $${}^1{\cal G}^{ab}_i\,=\, \sum_{j:j\neq i} b_{ji}^{ab}\, - \!\!   \sum_{k,j:  k>j>i} \!\! (J^{ab})_{cd}\, \ell^c_{kji} x^d_{kji}.$$}
\be
 \text{1-Gauss: }\q
{}^1{\cal G}^{ab}_i\,=\, \sum_{j:j\neq i} b_{ji}^{ab}\, - 2 \!\!   \sum_{k,j:  k>j>i} \!\!  \ell^{[a}_{kji} x^{b]}_{kji} 
\label{1GaussDb}
\ee
and the 1-Lorentz transformations it generates read, for $g_i\in\SO(4)$:
\be
 \label{eq:1Lorentz}
 \text{1-Lorentz: }\q
\begin{dcases}
h_{ji} \mapsto g_j^{-1} h_{ji} g_i \\
b_{ji} \mapsto g_i^{-1} b_{ji} g_i \\
\ell_{mlk}  \mapsto g_k^{-1} \act \ell_{mlk}\\
x_{mlk} \mapsto g_k^{-1} \act x_{mlk} 
\end{dcases}
\qquad (k<l<m).
\ee 

We remark that the form of the 1-Gauss constraint is somewhat non-symmetric, i.e. its expression depends on the node at which the constraint acts.\footnote{To reach a more symmetric formulation one could start with an extended phase space, where,  for a given dual face $f$, one defines three a priori independent variables $x_{f,n}$, with each of these rooted in a different node $n$ of the dual face. An alternative strategy is to consider a reduced phase space, where the 1-flatness constraints are partially implemented, and thus the choices for parallel transport should matter less. In fact, imposing the edge simplicity constraints constitutes a first step in this direction.}  E.g. the second sum in \eqref{1GaussDb} contains six terms at $i=1$, but no terms at $i=5$. The reason for this goes back to the way we defined the roots for the 2-holonomies $x_{kji}$, namely to be given by its {\it smallest} label $i$, $i<j<k$. As a consequence no 2-holonomy is based at $i=5$ while six of them are based at $i=1$.

\paragraph*{2-Gauss for 2-Lorentz}
The 2-Gauss constraint \eqref{2-GaussC} generates 2-Lorentz transformations \eqref{eq:12Lorentz}. 
These are parametrized by $\mathbb R^4$-valued 1-form $\eta$.
Upon discretization, these become transformations associated to the \textit{links} of $\Gamma$ with parameter $\eta_{ji}$ ($i<j$) defined in the frame of $i$, and similarly the 2-Gauss constraint is a $\mathbb R^4$ valued quantity  associated to the link $[ji]$ based at $i$. 
As discussed at length in the previous sections, the discretized 2-Gauss constraints takes the form of triangle closures:
\be
\text{2-Gauss}\q
{}^2{\cal G}_{\bar{kji}} = 
\begin{dcases}
-\ell_{\b{ki}} + \ell_{\b{kj}} + \ell_{\b{ji}} & \q\text{if } \bar i, \bar j, \bar k\neq \bar 1\\
-\ell_{\b{k1}} + h_{21}\act\ell_{\b{kj}} + \ell_{\b{j1}} & \q\text{if } \bar i= \bar1;\; \bar j, \bar k \neq \bar 2\\
-h_{32}\act\ell_{\b{k1}} + h_{31}\act\ell_{\b{k2}} + \ell_{\b{21}} & \q\text{if } (\bar i,\bar j)=(\bar 1,\bar 2);\; \bar k\neq \bar 3\\
-h_{42}\act\ell_{\b{31}} + h_{41}\act\ell_{\b{32}} + h_{43}\act\ell_{\b{21}} & \q\text{if } (\bar i,\bar j, \bar k)=(\bar 1,\bar 2, \bar 3) \\
\end{dcases}
\label{eq:triangle2}
\ee
The discrete 2-Lorentz transformation it generates then reads\footnote{Alternatively the action on $b_{ji}$ can be written as $$b_{ji}^{ab} \,\mapsto\, b_{ji}^{ab}  -  \sum_{k:k >j}  (h_{ji}J^{ab})_{cd} \,\eta_{kj}^c \ell^{\,d}_{kji} .$$}
\be
 \label{eq:2Lorentz}
 \text{2-Lorentz: }\q
\begin{dcases}
h_{ji} \mapsto h_{ji}\\
b_{ji}^{ab} \,\mapsto\, b_{ji}^{ab}  -   2{\sum}_{k:k >j}  (h_{ij}\act\eta_{kj})^{[a}_{\phantom{kj}}\!\ell^{\,b]}_{kji}  \\
\ell_{kji} \mapsto \ell_{kji}\\
x_{kji} \mapsto  x_{kji}  +  \eta_{ji} - \eta_{ki} + h_{ij}\act\eta_{kj}   
\end{dcases}
\qquad (i<j<k).
\ee
This action is also non--symmetric, e.g.~whereas we have for $b_{21}$ three terms appearing in the sum in (\ref{eq:2Lorentz}), there are no such terms for $b_{51}$. The reason for this lack in symmetry is again found in the need to choose explicit roots.

\paragraph*{1-flatness for 1-shift}
As usual from BF theory, the 1-flatness constraint is given by (the components of)
\be
 {}^1\!{\cal F}_{kji}\,=\, h_{ik}h_{kj}h_{ji} - \mathbb{1}_{\SO(4)} \q\q \q (i<k<j) .
\ee
This constraint sits at a face of $\Gamma$, and therefore it is associated to an edge of $\Delta_3=\pp\sigma$. Each edge is shared by 3 triangles which carry bivector variables conjugate to the three 1-holonomies involved in this constraint.
So, the Poisson flow of ${}^1\!{\cal F}_{kji}$ leaves the variables $\{h, \ell, x\}$ untouched, while it generates the discretized 1-shift symmetry for the three adjacent bivectors $b_{ji},b_{kj}$ and $b_{ik}$. 

{In formulas, for infinitesimal 1-shift parameters $\lambda_{kji} \in\so(4)$ and {\it on-shell} of the 1-flatness constraint itself,\footnotemark
\be
\text{1-shift:}\q
\begin{dcases}
h_{ji} \mapsto h_{ji}\\
b_{nm} \mapsto b_{nm} + \sum_{k:k>n>m} \lambda_{knm}  + \sum_{i:n>m>i} h_{mi} \lambda_{mni}^{a'b'} h_{mi}^{-1} - \sum_{j:n>j>m} \lambda_{njm}\\
\ell_{kji}\mapsto \ell_{kji}\\
x_{kji}\mapsto x_{kji}
\end{dcases}
\ee 
Notice that the right hand side above contains always 3 terms only: one per edge on the boundary of the triangle $[nm]^*$.
\footnotetext{%
This transformation property is obtained as follows. First the 1-flatness constraint is contracted with a group-valued parameter $\Lambda_{kji}\in\SO(4)$, to give the expression $\tr(\Lambda_{kji} {}^1\!{\cal F}_{kji} )$ which generates the following $h$-dependent translation of $b^{ab}_{nm}$ with $m<n$ (that we present {\it on-shell} of the 1-flatness constraint itself):
$$ 
b^{ab}_{nm} \stackrel{\Lambda_{kji}}{\mapsto} b^{ab}_{nm} + \delta^{nm}_{ji} \tr_{\SO(4)}(\Lambda_{kji} J^{ab})  + \delta^{nm}_{kj} \tr_{\SO(4)}( h_{ji}\Lambda_{kji} h_{ji}^{-1} J^{ab} ) - \delta^{nm}_{ki} \tr_{\SO(4)} (\Lambda_{kji} J^{ab})
$$
Now taking $\Lambda_{kji}={\bb 1}_{\SO(4)} + \lambda_{kji} + \dots$ for $\lambda_{kji} \in\so(4)$, at first order in $\lambda_{kji}$ we obtain
$$
b^{ab}_{nm} \stackrel{\lambda_{kji}}{\mapsto}b^{ab}_{nm} + \delta^{nm}_{ji} \lambda_{kji}^{ab}  + \delta^{nm}_{kj} (h_{ji})^a{}_{a'}(h_{ji})^b{}_{b'}\lambda_{kji}^{a'b'} - \delta^{nm}_{ki} \lambda_{kji}^{ab}.
$$
From which the equation in the main text is obtain by contraction with $J_{ab}$ and summation over the labels $i<j<k$.}
This is clear when the 1-shift transformation is expressed directly on the boundary triangulation:
\be
\text{1-shift :}\q
b_{\bar{kji}} \mapsto
\begin{dcases}
b_{\bar{kji}} + \lambda_{\bar{ji}} + \lambda_{\bar{kj}} - \lambda_{\bar{ki}} & \q\text{if }\bar i, \bar j,\bar k \neq \bar 1\\
b_{\bar{kj1}} + \lambda_{\bar{j1}} + \Ad^h_{21} \lambda_{\bar{kj}} - \lambda_{\bar{k1}} & \q\text{if } \bar i = \bar 1; \; \bar j, \bar k \neq \bar 2\\
b_{\bar{k21}} + \lambda_{\bar{21}} + \Ad^h_{31} \lambda_{\bar{k2}}  - \Ad^h_{32}\lambda_{\bar{k1}}  & \q\text{if }(\bar i, \bar j) = (\bar 1, \bar 2)\; \bar k \neq \bar 3\\
 b_{\bar{321}} + \Ad^h_{43} \lambda_{\bar{21}}  + \Ad^h_{41} \lambda_{\bar{32}}  - \Ad^h_{42}\lambda_{\bar{31}}  
& \q\text{if }(\bar i, \bar j, \bar k) = (\bar 1, \bar 2, \bar 3)
\end{dcases}
\label{eq:b1shiftDelta}
\ee%
where $\Ad^h_{ji}$ stands for conjugation by $h_{ji}$, e.g. $\Ad^h_{21}\lambda_{\bar{34}} := h_{21} \lambda_{\bar{34}} h_{21}^{-1}$.
This action is therefore also referred to as \textit{edge translations}.

Notice that this matches the continuum expression \eqref{eq:12shift}, once 1-flatness and the Stokes theorem are used.
All other canonical variables are left untouched by the flow of the 1-flatness constraint, cf. \eqref{eq:12shift}.}

\begin{figure}[t!] 
\centering
\begin{tikzpicture}[scale= 0.6 ,arrow1/.style={thick,->,shorten >=2pt,shorten <=2pt,>=stealth},
      arrow2/.style={thick,double distance = 3pt,-implies,shorten >=2pt,shorten <=2pt}  ] 
\begin{scope} 
\draw[thick] (0,0)--(3,0)-- (1.5,2.6) --cycle ;
\draw[arrow1] (0.55,0.4)--(1.5,2.05)--(2.45,0.4); 
\draw[arrow1]  (0.65,0.3) --(2.6,0.3) ;
\draw[arrow2](1.5,0.4)--(1.5,1.5) ;

\node[ left] at (0,-0.1) {$i$};
\node[right] at (3,-0.1) {$j$};
\node[above] at (1.5,2.7) {$k$};

\filldraw (0.49,0.24) rectangle (0.62,0.4);

\end{scope}

\begin{scope}[rotate = 60, xshift =0cm , yshift=0.2cm]  
\draw[thick] (0,0)--(3,0)-- (1.5,2.6) --cycle ;
\draw[arrow1] (0.55,0.4)--(1.5,2.05)--(2.45,0.4); 
\draw[arrow1]  (0.65,0.3) --(2.6,0.3) ;
\draw[arrow2](1.5,0.4)--(1.5,1.5) ;

\node[ left] at (1.5,2.6) {$m$};

\filldraw (0.55,0.3) circle (2.2pt);
\end{scope}

\begin{scope}[xshift =3.2cm , yshift=0.1cm,rotate = 60] 
\draw[thick] (0,0)--(3,0)-- (1.5,2.6) --cycle ;
\draw[arrow1] (2.45,0.4)--(1.5,2.05)--(0.55,0.4); 
\draw[arrow1]  (2.4,0.3)--(0.55,0.3)  ;
\draw[arrow2] (1.5,1.5)--(1.5,0.4) ;

\node[right] at (3,0) {$m$};
\filldraw (2.5,0.3) circle (2.2pt);
\end{scope}

\begin{scope}[rotate = 180, xshift = -3cm, yshift = 0.2cm] 
\draw[thick] (0,0)--(3,0)-- (1.5,2.6) --cycle ;
\draw[arrow1] (2.45,0.4)--(1.5,2.05)--(0.55,0.4); 
\draw[arrow1] (2.4,0.3) --(0.5,0.3);
\draw[arrow2] (1.5,1.5) --(1.5,0.4);

\node[ below] at (1.5,2.6) {$m$};
\filldraw (2.5,0.3) circle (2.2pt);
\end{scope}

\end{tikzpicture}
\caption{Definition of the $2$-flatness constraint.}\label{comp}
\end{figure}
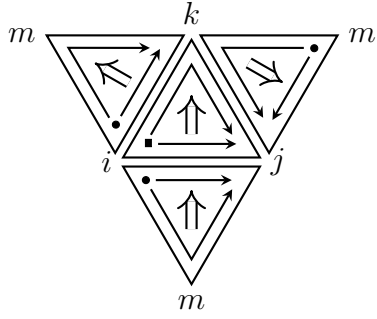

\paragraph*{2-flatness for 2-shift}
In the continuum, the 2-flatness constraint demands that the curvature of the 2-connection is vanishing. This curvature is a 3-form \eqref{eq:2-curvG}.
In the discrete, the 2-connection gives rise to 2-holonomies associated to the dual faces $[kji]$. We can compose the 2-holonomies associated to the dual faces around a dual 3-cell or bubble $[mkji]$ (with $i<j<k<m$), see figure \ref{comp}. The 2-flatness states that this composition vanishes:
\be
 {}^2\!{\cal F}_{mkji}\,=\, x_{kji} + x_{mki}-x_{mji}- h_{ji}^{-1}\act x_{mkj} \q \q (i<j<k<m) .
\label{2FlatD}
\ee
Note that we need to use the horizontal inverse for one face and the vertical inverse for another face. We have also chosen as root for the bubble the node $i$, as well as an explicit convention for the parallel transport for the dual face $[mkj]$ to $i$. 

The 2-flatness constraint  ${}^2\!{\cal F}_{mkji}$ generates the discrete analogue of the 2-shift symmetry \eqref{eq:12shift}. Its action on the edge vectors is nontrivial only for those four edge vectors that share the vertex $\bar{n}$ dual to the bubble $[mkji]$.
It also affects the $b$-variables, see \eqref{eq:12shift}. This again happens in a  non-symmetric fashion, as the constraint ${}^2\!{\cal F}_{mkji}$ (with $i<j<k<m$), associated to the vertex $\bar{n}=[mkji]^*$, only affects the $b$-bivector conjugate to $h_{ji}$, namely $b_{ji}$, see \eqref{2FlatD}. 
The variables $\{h,x\}$ are instead left untouched.

That is, for a 2-shift parameters $\mu_{\bar n = [mkji]^*}$ associated to the vertices of the triangulation (explicitly $\mu_{\bar n}= \epsilon_{\bar{n}mkji}\mu_{\bar n}$), we have the following action
\be
\text{2-shift: }\q
\begin{dcases}
h_{ji}\mapsto h_{ji}\\
b^{ab}_{ji}\mapsto b^{ab}_{ji} +  2\mu^{[a}_{mkji} (h_{ji}^{-1}\act x)^{b]}_{mkj}   \\
x_{kji}\mapsto x_{kji}
\end{dcases}
\qquad (i<j<k<m)
\ee
and
\be
\text{2-shift: }\q
\ell_{\bar{pn}} \mapsto
\begin{dcases}
 \ell_{\bar{pn}} - \mu_{\bar{n}} + \mu_{\bar{p}} & \q\text{if } \bar{n}, \bar p \neq \bar{1}\\
\ell_{\bar{p1}} - \mu_{\bar{1}} + h_{21}\act\mu_{\bar{p}} & \q\text{if } \bar n = \bar 1;\; \bar{p} \neq \bar{2}\\
 \ell_{\bar{21}} - h_{32}\act\mu_{\bar{1}} + h_{31}\act\mu_{\bar{2}} & \q\text{if } (\bar n,\bar p) = (\bar 1, \bar 2)
\end{dcases}
\qquad (\bar n < \bar p).
\ee

From these transformation properties of the edge vectors, we see that -- modulo the necessary parallel transports -- the 2-shift symmetry geometrically represents a translation of the 4-simplex' vertex $\bar n$ by a vector $\mu_{\bar n}$. 

{In $(3+1)$d, this vertex translation symmetry is understood as an incarnation of the action of the diffeomorphism symmetry group in the discrete \cite{Rocek:1982fr, Dittrich:2008pw}. This symmetry is expected to be broken on solutions involving (inhomogeneous) curvature \cite{Bahr:2009ku} or torsion \cite{Asante:2018wqy}. However, the BFCG dynamics imposes flat connections and vanishing torsion, and hence vertex translation symmetry is appropriately realized.
In this sense,  we interpret the 2-flatness constraint as the generator of \textit{4d diffeomorphism} thus subsuming both the \textit{spatial-diffeomorphism and Hamiltonian constraints in the tetrad sector.} At this purpose, notice that, as we are considering a vertex in the triangulation, there is no canonical choice of normal to the boundary surface and therefore no canonical decomposition of a 4-vector into a tangent and normal part to the discrete boundary.\footnote{Contraction of the 2-flatness constraints with $\ell$-dependent vectors will render their algebra non-Abelian. Nevertheless it will still be first class.  Due to the geometric interpretation of the constraints as generating vertex translations, the resulting algebra will be a discrete instantiation of the hypersurface deformation algebra. Regarding the algebra between constraints based at different vertices, we expect that this interpretation will be only fully consistent, if the 1-flatness constraints and triangle closure constraints hold. See \cite{Bonzom:2013tna} for a 3d analogue.}

Finally, it is important to emphasize that 2-shift symmetry only affects the $\ell$- and $b$-variables, but not the 1- and 2-connections. These are expected to get involved in the action of this symmetry (as required by the diffeomorphism constraint) if further (simplicity) constraints relating the $\ell$- and the $b$-variables are imposed. Such a step would be necessary for obtaining gravity, as opposed to a theory of flat space.}

\paragraph*{Discrete constraint algebra}

For completeness, we also include the non-vanishing elements of the constraint algebra. 
As expected, the constraint algebra is first-class.

The first four brackets simply state how the 4 constraints rotate under 1-Lorentz transformations:  
\begin{subequations}
\begin{alignat}{2}
\{ {}^1\!\cG^A_i, {}^1\!\cG^B_j\} &= \delta_{ij}\, {f^{AB}}_C \,\, {}^1\!{\cal G}^C_i,
&\qquad \{ {}^1\!{\cal G}^A_i, {}^1\!{\cal F}_{kji}\} &= {}^1\!{\cal F}_{kji}J^A-J^A\,\,  {}^1\!{\cal F}_{kji},\nonumber\\
\{  {}^1\!{\cal G} ^A_k , {}^2\!{\cal G}_{ij} \} &= - \delta_{kj} J^A\,\, {}^2\!{\cal G}_{ij}, 
&\qquad \{ {}^1\!{\cal G}^A_i, {}^2\!{\cal F}_{mkji}\} &= -J^A\,\, {}^2\!{\cal F}_{mkji}, 
 \end{alignat}
where for brevity we introduced the super-index $A=[ab]$ for labeling the generators of $\so(4)$, as in $b = \frac12 b_{ab}J^{ab} \equiv b_A J^A$, {and left matrix multiplications in the fundamental representation of $\SO(4)$ and its Lie algebra implicit};  the last nontrivial bracket is
 \begin{align}
 \{ {}^2\!{\cal G}^a_{nl}, {}^2 {\cal F}^b_{mkji}\} &= \delta_{nm}\delta_{lk}
(h_{ki} \,{}^1\!{\cal F}_{kji})^{ab}
& \qquad (i<j<k<m ; \; l<n).
\end{align}
 \label{disc-alg}
 \end{subequations}

\subsubsection{Which constraints to reconstruct the boundary geometry?}\label{recons}

As remarked above, the 2-Gauss constraint \eqref{eq:triangle2} suggests that we interpret the $\ell_{\b{ji}}$ variables as the edge vectors of a geometric triangles -- hence their namesake: triangle closures.
Moreover, the last four closures, all based at tetrahedron $\tau=[1]$, can be combined together to define a geometric tetrahedron, and not just triangles. 

However, not all the other tetrahedra, $\tau\neq[1]$ can be readily reconstructed in the same way from \eqref{eq:triangle2}, because not all triangle closures are based at the same tetrahedron. 

For example, consider the four triangle closures defining the tetrahedron $\tau=[2]$. By parallel transporting with $h_{21}$ the last of the closure equations of \eqref{eq:triangle2}, these can be successfully re-expressed as
\begin{align}
h_{21}\act\ell_{\b{45}} + h_{21}\act\ell_{\b{53}} + h_{21}\act\ell_{\b{34}} = 0; \quad&\quad
h_{21}\act\ell_{\b{34}} + \ell_{\b{41}} +\ell_{\b{13}} = 0; \\ 
h_{21}\act\ell_{\b{53}} + \ell_{\b{15}} + \ell_{\b{31}} = 0; \quad&\quad 
h_{21}\act\ell_{\b{45}} + \ell_{\b{51}} + \ell_{\b{14}} = 0; 
\end{align}
which shows that $\{h_{21}\act\ell_{\b{45}}, h_{21}\act\ell_{\b{53}}, h_{21}\act\ell_{\b{34}}, \ell_{\b{51}}, \ell_{\b{41}}, \ell_{\b{31}}\}$ consistently define tetrahedron $\tau=[2]$. 

On the other hand, the closures that are supposed to define $\tau=[5]$ read
\begin{align}
h_{43}\act\ell_{\b{12}} + h_{41}\act\ell_{\b{23}} + h_{42}\act\ell_{\b{31}} = 0;  \quad&\quad
h_{31}\act\ell_{\b{42}} + h_{32}\act\ell_{\b{14}} +\ell_{\b{21}} = 0; \\ 
h_{21}\act\ell_{\b{34}} + \ell_{\b{41}} +\ell_{\b{13}} = 0; \quad&\quad  
\ell_{\b{43}} + \ell_{\b{24}} + \ell_{\b{32}} = 0; 
\end{align}
and these cannot be written in such a way to involve only 6 edges vectors, as in the cases of $\tau=[1]$ or $[2]$ above.

In order to have five well-defined tetrahedra, it is sufficient to impose  the edge simplicity constraints (\ref{eq:edgesimpli1}), i.e. that for $i<j<k$
\be
\text{edge simplicity:}\quad
h_{ik}h_{kj}h_{ji} \act \ell_{kji} = \ell_{kji} \q .
\label{eq:edgesimplicity2}
\ee
This does also  imply, e.g. $h_{ji}h_{ik}h_{kj}\act (h_{ji} \act \ell_{kji}) =(h_{ji}\act \ell_{kji})$, that is the edge vectors are stabilized by the edge holonomies in the reference frames of each of the adjacent tetrahedra. Thus we can define $\ell_{kji}[j]= h_{ji} \act \ell_{kji}= h_{jk}h_{ki} \act \ell_{kji}$, without the need to specify in which direction along the face $[kji]$ (dual to the edge $\ell_{kji}$) is parallel transported to the tetrahedra $[i]^*$, $[j]^*$ or $[k]^*$.
Using the edge simplicity constraint and adopting this notation, it is easy to see that the triangle closures can be re-written in such a way that the five 6-plets of edge vectors 
$
\tau_k := \{\ell_{\bar{ij}}[k] \,:\, \b{i},\b{j} \neq k\}
$
define five tetrahedra, each in its own reference frame.

The edge simplicity constraints alleviate therefore some of the asymmetries that we encountered in section \ref{sec:constraints}, and that came from having to choose roots (and parallel transports) for the $x$ and $\ell$ variables. 

Further imposition of the 1- and 2-Gauss constraints completely resolves the asymmetry problem, but we will not show this fact in detail.

We conclude with a brief analysis of edge simplicity as a symmetry-generating constraint:

{
\paragraph*{Edge simplicity} 
The corresponding (transposed) constraint reads: 
\be
({\cal E}_{kji})_b\,:=\,(\ell_{kji})_a ( {}^1\!{\cal F}_{kji})^a{}_{b}  \,=\, (\ell_{kji})_a(h_{ik}h_{kj}h_{ji})^a{}_{b}  -  (\ell_{kji})_b\q,\q \q (i<k<j) .
\ee
We notice that 1-flatness has 6 independent components, while edge simplicity has 3 (the norm of $\ell$ is automatically conserved). Therefore, in this sense, edge simplicity is precisely {\it half} of the 1-flatness constraint (from the 1-holonomy perspective). However, as emphasized by the discussion above and that of section \ref{sec:contedgesimplicity}  -- where the continuum version of the edge simplicity constraint is shown to be {\it implied} by the continuum 2-version of the Gauss constraint -- we prefer to think of edge simplicity as complementing the discrete version of the 2-Gauss constraint. 

The flow of the edge simplicity constraint leaves untouched the edge vectors $\ell$ and the 1-holonomies $h$, while it transforms the bivectors $b$ and the 2-holonomies $x$. We shall focus on the action of parameter $v^b_{kji}$. Then, on the one hand, the bivectors transform  (on-shell of the edge simplicity constraint) in the same way as in a 1-shift transformation \eqref{eq:b1shiftDelta} whose parameter is the simple\footnote{A bivector $\beta \in \bb R^4 \wedge \bb R^4$ is said to be ``simple'' if it is of the form $\beta = v_1 \wedge v_2$ for some $v_1,\, v_2\in\bb R^4$.} bivector 
\be
\lambda^{ab}_{kji} = \ell_{kji}^{[a} v^{b]}_{kji}.
\label{eq:edgesimplbivector}
\ee 
On the other hand, the 2-holonomies get translated by a quantity proportional to the 1-flatness constraint: 
\be
x_{kji}^a\mapsto x_{kji}^a + ( {}^1\!{\cal F}_{kji})^a{}_{b}v^b_{kji}.
\ee
From (\ref{eq:edgesimplbivector}) we see that the only invariant information left in the bivector associated to the triangle $t=[kji]$ is the coefficient in front of  $\star (\ell_{kj} \wedge \ell_{ji})$ -- hence making the reduced phase space associated to the $(b,h)$ sector 2-dimensional and -- as we will make explicit below -- equal to $\mathrm T^*\mathrm{U}(1)$. This can be seen by expanding the bivector in an orthogonal basis that includes $\star (\ell_{kj} \wedge \ell_{ji})$, and observing that all other coefficients are subject to a translation under the action \eqref{eq:edgesimplbivector}.

Notice that edge simplicity Poisson-commutes with itself and with the 2-Gauss constraint, while it appropriately rotates under the action of the 1-Gauss constraint. This leads to a first class subalgebra of the complete constraint algebra analyzed above. This subalgebra of constraints can therefore be imposed to give rise to a reduced phase space. In the next two sections we will study the quantum version of this reduced phase space, that is the Hilbert space of G-networks. To build this Hilbert space, we will impose the 1- and 2-Gauss constraints as well as edge-simplicity directly at the quantum level. 

The remaining constraints, i.e. 2-flatness and the other half of the 1-flatness constraint (that, we recall, fixes the extrinsic geometry of the 4-simplex) will rather be understood as implemented by the ``dynamics'' of the model.


\subsubsection{Quantization of the constraints}\label{sec:quant}

Using standard techniques for the quantization of co-tangential bundles of compact or Abelian Lie groups, the constraints can be straightforwardly quantized.   
That is we adopt the Schr\"odinger representation of our discrete commutation relation on the Hilbert space
\be
{\cal H}= \mathcal L^2 \left( \bigotimes_{i<j} \SO(4)  \, \otimes \, \bigotimes_{i<j<k} \mathbb{R}^4 \right) \ni \tilde \Psi( h_{ji}, \ell_{kji}),
\ee
equipped with the $\SO(4)$-Haar and Lebesgue measures for the various factors. 
This is in analogy with the 3d construction, see \eqref{eq:64hilbert}.

Then, the matrix elements  $(h_{ji})_{ab}$ of the 1-holonomies and the vector components $\ell^a_{kji}$ of the edge vectors act as multiplication operators. 

Conversely, the 2-connection components act as derivatives
\be
x_{kji}^a \rightarrow  - i  \frac{\partial}{\partial \ell^a_{kji}}.
\ee
and the bivector fields $b^{a\bar{a}}_{ji}$ become left-invariant Lie derivatives $i L_{ji}^{a\bar{a}}$ on the variables $h_{ji}$:
\be
L_{j_oi_o}^{ab}\, \tilde \Psi (h_{ji}, \ell_{kji}) \,:=\, \frac{\d}{\d t}_{|t=0} \tilde\Psi( h_{j_oi_o} e^{ t J^{ab}},h_{ji \neq i_oj_o}, \ell_{kji}).
\ee

Quantization is now  straightforward, since the constraints (with phase space independent descriptors) are free of ordering ambiguities.\footnote{Note that the  $\ell^a x^b$ combination appearing in the 1-Gauss constraints is anti-symmetric in $(ab)$.} 

Let us consider some examples: The 2-Gauss constraints are multiplication operators, e.g. 
\be
{}^2{\cal G}^a_{52}\, \tilde\Psi \,=\, (- {(h_{21})^a}_b \ell_{521}+\ell^a_{532}+\ell^a_{542}) \tilde\Psi \q .
\ee
The 2-flatness constraints involve derivatives, e.g.
\be
{}^2\!{\cal F}^a_{4321}\, \tilde\Psi \,=\, -i \left( \frac{\partial}{\partial\ell^a_{321}} + \frac{\partial}{\partial\ell^a_{431}} - \frac{\partial}{\partial\ell^a_{421}} - (h_{ij})_{ab} \frac{\partial}{\partial\ell^b_{432}}\right) \tilde\Psi ,
\ee
while the 1-flatness constraint acts by multiplication
\be
({}^1\!{\cal F}_{321})_{ab} \,\tilde \Psi \,=\, \Big( ( h_{13}h_{32}h_{21})_{ab} - \delta_{ab}\Big)\tilde \Psi,
\ee 
and so does the the edge simplicity constraint
\be
({\cal E}_{321})_b \, \,\tilde \Psi \,=\,  \Big( \ell^a_{321}( h_{13}h_{32}h_{21})_{ab} - (\ell_{321})_b\Big) \,\tilde \Psi .
\ee
Finally, the 1-Gauss constraints is
\be
{}^1\!{\cal G}^{ab}_{i}\, \tilde\Psi \, \,=\,  i \left(\sum_{j \neq i} L_{ji}^{ab} \, - \sum_{k>j>i}(J^{ab})_{cd} \ell^c_{kji} \frac{\partial}{\partial \ell_{kji}^d}\right) \tilde\Psi   .
\ee

Given the boundary data  $(h_{ji}, \ell_{kji})$ associated to $\Gamma=(\pp \sigma)^*$, the next step is to find how to parametrize the solutions of  the 1- and 2-Gauss constraints  $^2\cG$, together with the edge simplicity constraints. We will call these states {\it G-networks}.

\subsubsection{G-networks}\label{sec:gnetwork}

Based on the results of the sections \ref{sec:getu1} and \ref{recons}, which provided the geometric meaning of the imposition of   edge simplicity constraints  as well as the 1- and 2-Gauss constraints, we expect to reduce the boundary data $(h_t, \ell_e)\in \SO(4)\times \mathbb{R}^4$ to the KBF boundary data $(s_t, l_e)\in (\mathbb{Z}, \mathbb{R}^+)$, where the integers $s_t$ are associated to the triangles and the edge lengths $l_e$ are associated to the edges.

The reduction follows directly from the previous discussions.  We have argued in section \ref{recons}, that we need to impose the 2-Gauss constraints (aka triangle closure constraints) and the edge simplicity constraints, in order to be able to reconstruct the geometry of the tetrahedra. On the other hand, we have seen in section \ref{sec:getu1} that the edge simplicity constraint allow to restrict $h_t$ to one free parameter $\theta_t$ per  triangle (or link) as free parameter in the holonomies $h_t$. Hence they allow to reduce $\SO(4)$ to $\textrm{U}(1)$.  A $\text{U}(1)$ Fourier transformation leads to conjugated variables $s_t \in \mathbb{Z}$.

The edge lengths, $l_e=\sqrt{\ell_e\cdot \ell_e}$, arise from the edge vectors if we also impose 1-gauge invariance, that is the 1-Gauss constraints, just like for the 3d BF case.   Note that the dihedral angles $\theta_t$, as defined in (\ref{thetadef0}) are also 1-gauge invariant, as they arise from the $\SO(4)$--invariant inner product between two normal vectors.

The question is now whether the edge lengths $l_e$ and the $\theta_t$ are sufficient to reconstruct the data $\ell_e$ and $h_t$. To this end consider a triple $e \subset t \subset \tau$ and first assume that the frame for the tetrahedron is put into extended time gauge. That is, we choose the edge vectors of the tetrahedron to be all orthogonal to $(1,0,0,0)$, and also demand that the edge vector associated to $e$ is parallel to $(0,1,0,0)$ and that the edge vectors of the triangle $t$ lie in the plane spanned by $(0,1,0,0)$ and $(0,0,1,0)$. This together with the condition that the tetrahedron lies in the positive quadrant completely fixes the 4d rotational freedom, and shows that we can reconstruct from the six edge lengths the edge vectors $\ell_e[\tau]$. Assuming that also the other tetrahedron $\tau'$ sharing $t$ is in extended time gauge, we can, by including $\theta_t$, reconstruct $h_t$ and the edge vectors $\ell_e[\tau']$ in this tetrahedron. One can continue this process by changing the tetrahedral frames to extended time gauges with respect to other choices of triples, in order to reach other neighbouring tetrahedra.

Thus, defining $G$-network functions as wave functions being annihilated by the 1-Gauss, 2-Gauss and edge simplicity constraints, 
\begin{eqnarray}\label{eq:ConstraintsForG}
{}^1{\cal G}_k  \tilde \Psi =0 \q ,\q\q {}^2{\cal G}_{ji}  \tilde \Psi =0\q ,\q\q {\cal E}_{kji} \tilde \Psi =0 \q ,
\end{eqnarray}
the G-network functions should be vanishing for $(h_t,\ell_e)$ configurations, which do not satisfy the 2-Gauss constraints or the edge simplicity constraints,  and only depend on the $((h_t,\ell_e)$ via the $(\theta_t,l_e)$ data, that is $\tilde \Psi( h_t, \ell_e)=\tilde \Psi(h_t(\theta_t,l_e), \ell_e(\theta_t,l_e))=:\tilde \Psi(\theta_t,l_e)$.

Thus, a basis of 4-simplex G-networks is labelled by 10 edge lengths $l_e\in\mathbb R_+$ and 10 angles $\theta_t\in\mathrm{U}(1)$ that can be Fourier transformed to give 10 integers $s_t\in\mathbb Z$. These, $\{(l_e, s_t)\}$, are precisely the data that label the irreducible 2-representations of the Poincar\'e 2-group -- see section \ref{sec:KBF}.

One can construct an inner product on this space of wave functions $\tilde \Psi(\theta_t,l_e)$, by using one of the methods discussed in \cite{Giulini:1998rk, Ashtekar:1994kv, Dittrich:2004bn, Louko:2005qj}, which propose and apply a variety of techniques to obtain induced inner products on constraint solution spaces. Here the 1-Gauss constraints and the 2-Gauss constraints are less problematic, as these correspond to compact gauge groups and an Abelian (but non-compact) gauge group, respectively. The edge simplicity constraints are however more involved, also due to complicated redundancies between the delta functions imposing these constraints. We therefore leave this problem to future work.

\subsection{G-networks versus Poincar\'e networks}\label{sec:bfcg-BF}

It has been pointed out that the BFCG action can be seen as a Poincar\'e BF theory -- up to a boundary term -- when dealing with the  Poincar\'e (2-)group \cite{Mikovic:2011si,Mikovic:2018vku}. 

\begin{eqnarray}\label{eq:BFCG-BF}
S_{BFCG} &=& \int_{M_4} \tr_{\so(4)}( B \wedge F[A]) + \tr_{\mathbb R^4}(C \wedge G[\Sigma, A] )\nonumber\\
&=& \int_{M_4} \tr_{\so(4)}( B \wedge F[A]) - \tr_{\mathbb R^4}(\Sigma  \wedge \rd_A  C ) \nonumber\\
&=& \int_{M_4} \tr_\text{Poinc}( \cB \wedge \cF[\cA]) - \int_{\partial M_4} \tr_{\mathbb R^4} (\Sigma  \wedge  C ),
\end{eqnarray}
 where we used the Poincar\'e Lie algebra valued connection $\cA=A+C$,  its associated curvature $\cF(\cA)= F(A) + \rd_AC$, and the 2-form $\cB=B+\Sigma$ with value in the Poincar\'e Lie algebra.  We also use the natural Killing form $\tr_\text{Poinc}$ for the Poincar\'e Lie algebra.  
 
At the continuum level, the two actions can be seen as two different choices of polarization in the translation sector. 
However, while at the continuum level the choice of polarization does not matter, upon discretization, it does. 

This was for example emphasized in 3d gravity \cite{Dupuis:2019unm}, where two different discretizations (with each polarization) are recovered from the  two possible continuum polarization choices which lead to different vacua \cite{Dittrich:2014wpa, Bahr:2015bra, Delcamp:2016yix, Delcamp:2018sef, Delcamp:2018efi}. One should expect nevertheless that the different discrete pictures are related. 
 In 3d this provided an interesting duality between two classes of  representation of the Drinfeld double \cite{Freidel:2006qv, Buerschaper:2010yf} (in the case of finite gauge group). In 4d, this might lead to uncovering new  dualities. A possible direction to follow is to compare our work with that of \cite{Freidel:2019ees} on Poincar\'e networks.

In the present case, the BFCG choice seems to be more adequate than the $\cal BF$ one to describe quantum simplicial geometries in 4d. Indeed, \textit{on the boundary graph }$\Gamma$, the natural discretization of the $\cB\cF$ theory would lead to a discretization of the Poincar\'e connection $\cA=A+C$ on the links of $\Gamma$, whereas the $\cB$ field would be discretized on the triangles (the dual of the links). As we emphasized at length, the $C\equiv E$ field encodes the tetrad. To recover the metric information of the discrete geometry, we therefore expect the tetrad to decorate the edges of the triangulation and not on the links of its dual. This is the main advantage of the 2-gauge theoretical perspective provided by  the BFCG-Yetter model.
    
\begin{table}[t]
\begin{center}
\begin{tabular}{c|c|c|c|}
\multicolumn{2}{c}{}& \multicolumn{1}{@{}c@{}}{$\overbrace{\hspace*{\dimexpr6\tabcolsep+2\arrayrulewidth}\hphantom{012}}^{\text{1-gauge th.}}$}&\multicolumn{1}{@{}c@{}}{$\overbrace{\hspace*{\dimexpr6\tabcolsep+2\arrayrulewidth}\hphantom{012}}^{\text{2-gauge th.}}$}\\
\cline{2-4}
&\cellcolor{black!75} \small$\color{white}[C\equiv E]$& $\cB\cF $    & BFCG                  \\ \cline{2-4}
\ldelim\{{2}{5.4mm}[\;$\Gamma$\;]  &links & $(A, E)$ & $A$\tikzmark{a}
\\\cline{2-4}
&faces & $-$ &  $\Sigma$\tikzmark{s} 
\\ \cline{2-4}
\ldelim\{{2}{8.2mm}[\;$\Delta_3$]&triangles & $(B,\Sigma)$ & $B$\tikzmark{b}
\\ \cline{2-4}
&edges &$-$ & $ E$\tikzmark{e} 
  \\ \cline{2-4}
\end{tabular}
\caption{We summarize where the different fields are discretized according to the choice of polarization of the continuum theory. We use the interpretation of $C$ as the tetrad field, $C\equiv E$, and the definition of the Poincar\'e-BF connection and bivector: $\mathcal A =(A, E)$ and $\mathcal B = (B,\Sigma)$. $\Delta_3$ is here the boundary triangulation of the 4-simplex $\sigma$, i.e. $\Delta_3=\pp\sigma$, while $\Gamma$ is its dual $\Gamma = \Delta_3^*$. The arrows indicate pairs of conjugate variables as well as pair of dual cells in $\Gamma$ and $\Delta_3$.}\label{tab:BFCG-BF}
\end{center}
\begin{tikzpicture}[overlay, remember picture, yshift=.25\baselineskip, shorten >=.5pt, shorten <=.5pt, line width=.75pt]
    \draw [<->] ([xshift=10mm]{pic cs:a}) [bend left=72] to ([xshift=10mm]{pic cs:b});
    \draw [<->] ([xshift=10mm]{pic cs:s}) [bend left=72] to ([xshift=10mm]{pic cs:e});
\end{tikzpicture}
\end{table}

As we have already emphasized in section \ref{sec:constraints}, the BFCG polarization allows then to see the 2-flatness constraint as implementing a translation of the vertex, that is a discretized diffeomorphism. Instead in the $\cB\cF$ polarization, we can only have translations of edges, so that the nice geometric discretization of the diffeomorphisms is lost in the Poincar\'e BF picture. Another way to make this point is to emphasize (see section \ref{sec:BFCG-Yetter}) that whereas the boundary data of a 4d BF amplitude is labelled by a 2-complex, in the BFCG-Yetter model, the boundary data is labelled by the (dual) \textit{3}-complex $\Gamma$, which allows a better grasp of the geometry. 
From this perspective, the BFCG polarization seems to be better suited to discuss the discretization of gravity.

\section{Derivation of the KBF model from the BFCG-Yetter model}\label{KBFfromBFCG}
We have now the key ingredients to relate the KBF model to the BFCG-Yetter model. The edge simplicity together with the discretized 1- and 2- Gauss constraints have been used to define the boundary state $\Psi_\Gamma$. Just like in 3d, we expect the BFCG-Yetter amplitude to act as a projector on the 2-curvature constraint.

We  found that the partition function with a (simplex) boundary is given by (\ref{eq:discreteZ4}), which we here display again: 
\begin{align}\label{discreteZ4p}
&Z_Y(\sigma|\tilde\Psi) =\notag\\
&\int \DD h_{l} \DD \ell_{f} \!\! \!\! \prod_{f\in(\pp\sigma)^*}\!\!\!\delta_{\SO(4)}\!\bigg( \overleftarrow{\prod_{l\in\pp f}} h_{l}^{\epsilon(l|f)} \bigg)   \!\!
\prod_{l\in(\pp\sigma)^*}\!\!\delta_{\bb R^4}\! \bigg(\sum_{f^* \subset l^*} \!\!\!\epsilon(f^*|e^*) \ell'_{f}\bigg)
\tilde \Psi(h_{l},\ell_{f})\, .
\end{align}
We argued that imposing the 2-Gauss constraints, the edge simplicity constraints and the 1-Gauss constraints reduce the boundary wave functions to $\tilde \Psi(\theta_t,l_e)$. The 2-Gauss constraints (aka triangle closure constraints) appear explicitly in (\ref{discreteZ4p}), as do the edge simplicity constraints, which are part of the 1-flatness constraints. The projector on the 1-Gauss constraints is imposed implicitly, through the fact that the amplitude kernel is invariant under 1-gauge transformations, and therefore imposes a group averaging over these. Similarly, the 2-flatness constraints are imposed also by group averaging thanks to the invariance of the amplitude kernel under 2-shift symmetry.

Thus, among the constraints appearing explicitly in the delta-functions of  \eqref{discreteZ4p}, we are only left with the ``reduced'' 1-flatness constraints -- i.e. what is left of it after imposition of edge simplicity and triangle closure.  
We have seen in section \ref{sec:1flatness4LC}, that these  reduced constraints fix the angles $\theta_t$ to be the exterior 4d dihedral angles $\Theta_t(\{l\})$ between the tetrahedra when the 4-simplex is embedded in flat Euclidean space -- up to a global orientation $\epsilon$.
Thus, on shell of triangle closures and edge simplicity, the only constraint left to be explicitly imposed in the amplitude is given by the following $\mathrm U(1)$ delta-function:
\be
\sum_{\epsilon=\pm1} \prod_{t\in\pp\sigma} \delta_{\mathrm U(1)}( \exp(i (\theta_{t}- \epsilon \Theta_{t}(l))  )=\sum_{\epsilon=\pm1} \prod_{t\in\pp\sigma} \sum_{s_{t}\in \mathbb{Z}} \exp(-i s_{t}(\theta_{t}- \epsilon \Theta_{t}(l_e))) \quad .
\ee
Thus, on-shell of the triangle closure and edge simplicity, and in terms of the  $\mathrm U(1)$-Fourier transformed boundary state $\check \Psi$, defined by
\be
\tilde \Psi(\theta_t, l_e) \,=\,  \sum_{s_t \in \mathbb{Z}}  \prod_{t\in\pp\sigma}  \exp( i s_t \theta_t) \, \check \Psi(s_t,l_e),
\ee
the BFCG-Yetter partition function \eqref{discreteZ4p} can be written as
\begin{align}
Z_Y(\sigma|\check\Psi) =
\sum_{\epsilon=\pm1}\int  \DD l_e \DD s_t \, \mu(s_t,l_e) \, \frac{\exp(i \epsilon s_{t}\Theta_{t})}{4!\text{Vol}_\sigma(l_e)} \check \Psi(s_t, l_e).
\end{align}
Similarly to what we argued for the PR model, the appropriate bulk measure factor for the KBF model \eqref{eq:Z_KBF} can be fixed by the requirement that the resulting partition function is triangulation invariant (the original BFCG-Yetter partition function is) \cite{Korepanov:2002tp,Korepanov:2002tq,Korepanov:2002tr,Baratin:2006gy, Baratin:2014era}. The measure entering this boundary-state amplitude is instead more ambiguous, since it depends on normalization choices for the boundary states. If these are normalized, then the measure $\mu(s_t , l_e)$ is
\be
\mu(s_t,l_e)\,=\, \prod_t \sqrt{2A_t(l_e)}
\ee
With this observation, we reproduce the amplitude kernel for the KBF partition function and thus prove the sought equivalence between the BFCG-Yetter and KBF models:
\be
Z_Y(\sigma | \check \Psi) = Z_{KBF}(\sigma| \check\Psi)
\ee

In summary the KBF amplitude kernel arises from a Fourier transformation of the delta functions that impose the reduced 1-flatness constraints. Thus one can expect that the model features 1-shift symmetry. One might wonder what happened with the 2-shift symmetry, which in the Yetter model (\ref{discreteZ4p}) imposes the 2-flatness constraints. The work \cite{Baratin:2006gy} shows that the KBF model does indeed feature 1-shift and 2-shift symmetries, and provides a (Fadeev-Popov) gauge fixing procedure, to make the model well-defined.

\section{Discussion and outlook}\label{outlook}
\textit{``The journey matters more than the destination." } In this paper we wanted to prove the conjecture \cite{Baratin:2009za} that the KBF model is a state sum for the BFCG-Yetter model.
In doing so we actually uncovered several aspects of 4d discrete geometry which provide new insights on (3+1)d quantum geometry. These insights might reveal themselves as more important than the mere equivalence of the two models since they can lead to genuinely new developments in the model building of 4d quantum gravity.

\paragraph*{Avoiding the 2-Peter-Weyl theorem} 
 Since we lack the 2-Peter-Weyl theorem for 2-groups or the Poincar\'e 2-group in particular, the proof of the equivalence could not be done head on, as it can be done in 3d. Fourier transforming the BFCG-Yetter  partition function as discussed in \cite{Mikovic:2011si} does not work either.  One technical insight provided in this paper, is that to arrive at the KBF model, we have to \textit{solve} part of the delta--functions, which appear in the Yetter amplitude, explicitly, and apply the Fourier transform only to the remaining delta--functions. This reduces the partition function to a summation over a reduced configuration space: the original path integral over an general $\so(4)$-valued 1-connection and a tetrad field  (conjugate to the 2-connection) is reduced to a path integral over torsion-free configuration data as restricted by the imposition of the 2-Gauss  and edge simplicity constraints.  The torsion freeness condition turns out to restrict (in the discrete) both the tetrad variables and the connection. Imposing 1-gauge invariance of the torsion free data leads to 2-group representation labels.

We came to this insight by introducing a boundary and boundary states $\Psi_{\Gamma}$,  which provided us  guidance on which constraints we had to impose in order to identify the KBF quantum state within the more general BFCG-Yetter Hilbert space.  We found that the relevant ``kinematical'' space is a solution of the 1- and 2-Gauss constraint -- as we could have naively expected -- but also of the edge simplicity constraints. 
Technically, these are a subset of the 1-flatness constraints; however, they are best understood from a geometrical perspective as being part of a correct discretization of the {\it continuum} 2-Gauss constraint. This is in fact equivalent to a torsionfreeness condition for the tetrad field $C\equiv E$ -- a condition that cannot be solely imposed by the {\it discrete} version of the 2-Gauss constraint, which is merely given by triangle closures (see sections \ref{sec:contedgesimplicity}, \ref{sec:boundaryedgesimplicity}, \ref{sec:getu1}, and \ref{recons}).

\paragraph*{Edge simplicity in (loop) quantum gravity} 
A natural expectation in loop quantum gravity (LQG) is that a simplicial version of the LQG Hilbert space would describe, after quotienting out the $\SU(2)$ gauge symmetry,  the degrees of freedom of simplicial gravity as given by the canonical formulation of Regge calculus \cite{Regge:1961,Dittrich:2009fb,Dittrich:2011ke}. 
However, this expectation is not correct: the LQG phase space turns out to describe additional degrees of freedom that encode the possibility for a given triangle to have different shapes prescribed by the geometrical data of the two neighbouring tetrahedra sharing it \cite{Dittrich:2008va, Dittrich:2008ar, Freidel:2010aq}. This fact has been argued to be related to the presence of torsion degrees of freedom \cite{Haggard:2012pm,  Asante:2018wqy}. 
Now, the difference between the LQG phase space and the Regge phase space can be shown to originate precisely in the (missed) imposition of a discrete version of the secondary constraints \cite{Dittrich:2010ey}. Let us recall quickly the origin of these  the secondary constraints. 
State sum models for 4d quantum gravity (aka spinfoam models) are characterized by the slogan ``first quantize and then constrain''. Based on a triangulation, they start from a discrete quantum partition function (state sum) for the topological BF theory for the Lorentz group and aim at imposing the simplicity constraints at the quantum level. The role of these constraints is to reduce the (quantum version of the) bivector field $B$ to be of the form $\star E\wedge E$ and thus reducing BF theory to quantum gravity.\footnote{The Immirzi parameter can also be included, if desired.}
The {\it primary} simplicity constraints then are equivalent to demanding that the variables $\star B_t$, encoding the discretized B field, associated to the triangles of a given tetrahedron span (in their internal space) only a three-dimensional space and hence are all orthogonal to some vector, which defines the normal to this tetrahedron\footnote{Notice that the primary simplicity constraints are related to the triangle closure constraints: these ensure that from the six edge vectors associated to a tetrahedron only three are independent, and thus allow to define a normal to this tetrahedron.} \cite{Engle:2007wy, Freidel:2007py}.
This condition can be translated into a restriction on the type of allowed representations in the state sum \cite{Barrett:1997gw, Freidel:2007py, Engle:2007wy}.
Nonetheless, the appearance of \textit{secondary} simplicity constraints in the canonical analysis of the primary simplicity constraints raises the question of whether the imposition of the primary simplicity constraints we just described is sufficient to reduce the state sum of BF theory to one for quantum gravity, and in particular whether the secondary simplicity constraints should also be imposed explicitly \cite{Alexandrov:2011ab, Bonzom:2009hw, Dittrich:2008va, Dittrich:2008ar, Anza:2014tea, Hellmann:2013gva, Oliveira:2017osu}.  The key point is that these  secondary constraints take precisely the form of the edge simplicity constraints \cite{Dittrich:2010ey, Dittrich:2012rj}, we have discussed here.

In the light of the relationships between twisted geometries and torsion, and between edge simplicity and torsionfreeness, this result should not be too surprising. It is however remarkable from the perspective of the present work that edge simplicity is so profoundly connected to the dynamical problems faced by the spinfoam approach to LQG.

\paragraph*{Discrete  diffeomorphisms}
 As we have just recalled, the standard  spinfoam models for LQG are built using as a backbone a BF theory for the Lorentz group, discretized on a 2-complex. As a consequence, a natural implementation of the discretized diffeomorphisms in the form of a vertex translation symmetry \cite{Dittrich:2011ke} is either missing or very cumbersome. 
Let us exemplify this by reviewing two seemingly natural constructions. 
Waelbroeck and Zapata \cite{Waelbroeck:1993sm} noticed that (in the boundary discretization) it is possible to break down the ``edge translations'' generated by the curvature 2-form $F$ smeared on the dual faces into the subgroup of vertex translation. To achieve this at a given vertex one could build a constraint generating the vertex' translations out of certain combinations of the curvatures associated to the faces bounding the 3-bubble dual to this vertex. However, the problem is that these constraints will inevitably affect the neighbouring vertices too, thus leading to highly nonlocal constructions if only one vertex at the time has to be affected.
Thiemann instead, proposed a discretization of the LQG Hamiltonian \cite{Thiemann:1996ay, Thiemann:1996aw} anchored at the nodes of the boundary spin-network graph, i.e. at the tetrahedra of the boundary triangulation.
However, as was pointed out by Immirzi \cite{Immirzi:1994gq}, this does not implement the natural geometric action of diffeomorphisms which is present in (linearized) Regge calculus \cite{Rocek:1982fr,Dittrich:2009fb}.

In contrast to these two examples, the BFCG-Yetter-KBF formulation provides the right framework to recover discretized diffeomorphisms: its key feature is that the tetrad field is present from the onset and is discretized on the triangulation edges. 
Conjugate to the tetrad field is the 2-connection $\Sigma$ which is discretized over the faces of $\Gamma$. The 2-connection leads to a 2-curvature, which is a 3-form, and is based on the bubbles of the dual complex. In a 3d discretization these are dual to vertices, and this allows the discrete 2-flatness constraints to generate vertex translations. We showed, in section \ref{sec:constraints}, that this is indeed the case.  Hence we can identify the 2-flatness constraints with  {\it the Hamiltonian and diffeomorphism constraints} of the Yetter model. 
Indeed, as in (linearized) Regge calculus \cite{Dittrich:2009fb}, these constraints are based at the vertices of the triangulation and give correctly four constraints per vertex.   
Note also that the form of the constraints is very simple. 
This fact can of course change, if we impose further geometricity constraints on the bivector field.

The (spatial) diffeomorphism constraints have been recently revisited in LQG through the introduction of corners (that is boundaries for the boundary hypersurface) and associated corner data into the framework. This leads to an extension of spin networks to bubble and Poincar\'e\footnote{More precisely, spin-networks for the 3d Euclidean $\mathrm{ISO}(3)$ group, due to the choice of time gauge.} networks \cite{Freidel:2019ees, Freidel:2018pvm}. The form of the  (spatial) diffeomorphism constraints in \cite{Freidel:2019ees} is very similar to the 2-flatness constraints, discussed in this work. 
However, it is important to notice that here we do not require the introduction of corners. Rather, in our case, the topological structures needed to support the point-like excitations are naturally provided by the categorification of the theory and in particular by the introduction of a 2-connection. Indeed, this categorification was precisely meant to allow for coupling of lower dimensional defects  \cite{2012arXiv1211.0529B}. In this sense, we believe that using 2-gauge theoretical structures provides a more fundamental approach that can {\it also} lead to a more systematic understanding of corners.

\paragraph*{Outlook}
There are several avenues to pursue in future studies.  First, a more precise definition of the Hilbert space of the G-networks needs to be given. In particular the measure/scalar product needs to be fully understood. Another, related, line of research concerns the understanding of the measure appearing in the KBF model. We have not derived it strictly speaking from the BFCG-Yetter model, but used general arguments of topological symmetry and previous results. An explicit, head-on, computation should however be possible. 
 More broadly, we also speculate that a good understanding of the G-networks Hilbert space should provide valuable clues to understand which form a 2-Peter-Weyl theorem has to take; the most promising route seems in fact to retro-engineering the ``Fourier transform'' used in our definition of the G-networks. 

Another avenue to explore concerns holography, which was actually among the first motivation to start this project. The meaning of holography can be explicitly checked using topological models, as was realized in 3d \cite{Bonzom:2015ans, Dittrich:2017hnl, Dittrich:2017rvb, Dittrich:2018xuk, Riello:2018anu,Asante:2019ndj}
and 4d \cite{Asante:2018kfo}. For example in 3d, one can show that the boundary theory, dual to the Ponzano-Regge model (with metric boundary condition)  is given by certain 2d spin-chain models. 
The BFCG-Yetter-KBF model is a 4d topological model which shares many geometric features with a possible theory of quantum gravity: it is thus of great interest to explore its dual boundary theories.

From the perspective of topological quantum field theory and topological invariants, one expects that in 4d  interesting topological invariants should be built from a categorified version of quantum groups \cite{Baez:2009as}. 
In 3d, quantum groups are associated to homogeneously curved quantum simplicial geometries \cite{Mizoguchi:1991hk, 2003math......5113T}. 
By analogy, we could expect that introducing curved simplicies in 4d would provide the right framework to construct such invariants. 
While a viable notion of categorified quantum group is still missing, now that we have a promising notion of discrete geometry (the one associated to the G-networks), we can try and bypass this roadblock in the same way that we bypassed the lack of the 2-Peter-Weyl theorem: i.e. by following the geometric intuition and ``curving'' the associated quantum geometries  \cite{Haggard:2014xoa, Haggard:2015kew, Riello:2017iti, Delcamp:2016lux, Dittrich:2017nmq}.

Finally our main interest being gravity, we should see how imposing the simplicity of the bivector field $B$ changes the picture \cite{Mikovic:2011si, Mikovic:2018vku}. Note that currently, at the topological level, the G-networks are given in terms of the tetrad field discretized on the edges of the triangulation, whereas the $B$ field is discretized on the triangles of the triangulation. The tetrad field and the $B$ field are independent to each other: from the point of view of gravity we have doubled the size of field space. Now, imposing the simplicity of the $B$ field by tying it to the tetrad field, will render the geometric picture fully consistent, i.e. by tying the bivector data at the triangle to the edge information. We expect that such constraints between bivector and tetrad fields lead to further (secondary) constraints, relating also the 1-connection with the 2-connection. As mentioned earlier, it would be interesting to see how the diffeomorphisms are realized in this more geometric picture. We leave this to future investigations.

\pagebreak

\begin{center}
\textbf{Acknowledgements }
\end{center}
We thank Sylvain Carrozza and  especially Ma\"it\'e Dupuis for suggesting and organizing the PI--Waterloo Quantum Gravity Winter Retreat, where work on this project was started. We thank Camp Kintail for providing the venue and a humungous amount of food. 
SKA is supported by an NSERC grant awarded to BD. PT is supported by an NSERC grant awarded to FG. This work is  supported  by  Perimeter  Institute  for  Theoretical  Physics.   Research  at  Perimeter  Institute is supported by the Government of Canada through Industry Canada and by the Province of Ontario through the Ministry of Research and Innovation.

\begin{appendix}
\section{From bulk to boundary amplitudes in the BFCG model}\label{app:1}

Here we expand on the discussion in section \ref{sec:BFCG-Yetter} and describe more explicitly how to compute the BFCG amplitude for a four-simplex with boundary.  Thus we start with the dual of the 4-simplex and the dual 3-comples $\Gamma$ to its boundary. The dual 4-simplex is represented in Fig. \ref{penta2} with a node labeled $0$ (the blue star), and the one--skeleton of the dual 3-complex is depicted in red, with its dual tetrahedra labelled by $i=1,\ldots, 5$. For the boundary triangulation and dual we will use the same conventions as in section \ref{sec:5}. 

 As shown in Fig. \ref{penta2} we have bulk links (dashed blue lines) $[i0]$, going from the node $0$, which represents the dual simplex, to the nodes $i$, which represent the nodes dual to the tetrahedra in the boundary. To these bulk links we associate bulk holonomies $H_{i0}$, which parallel transport from the node $0$ to $i$. The inverse holonomies are denoted by $H_{0i}$.
 
 We have furthermore 10 bulk faces $[ji0]$, with $j>i$ and $i,j, \ldots=1,\ldots 5$. To these bulk faces we associated 2-connection (bulk) variables $X_{ji0}$, rooted at $0$.

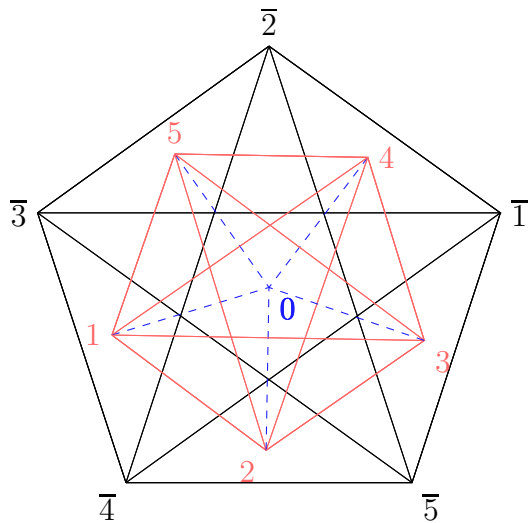
\begin{figure}[ht!]
\begin{center}
\begin{tikzpicture}[scale=0.8]%
\def\ra{4}
\def\rb{2.7}
\begin{scope}
\foreach \x in {18,90,...,306} {
   	\draw (\x:\ra) node{} -- (\x+72:\ra) ;
       	\foreach \y in {18,90,...,306}{
		\draw (\x:\ra) -- (\y:\ra) ;
		} ;
};
\node[right] at (18:\ra) {$\bar{1}$};
\node[above] at (90:\ra) {$\bar{2}$};
\node[left] at (90+72:\ra) {$\bar{3}$};
\node[below left] at (90+2*72:\ra) {$\bar{4}$};
\node[below right] at (90+3*72:\ra) {$\bar{5}$};
\end{scope}
\begin{scope}[rotate = 35,red!60]
\foreach \x in {18,90,...,306} {
   	\draw (\x:\rb) node{} -- (\x+72:\rb) ;
       	\foreach \y in {18,90,...,306}{
		\draw (\x:\rb) -- (\y:\rb) ;
		} ;
\draw[dashed,blue!90] (\x:\rb) -- (\x:0) node[below right]{0} ;
};
\node[right] at (18:\rb) {4};
\node[above] at (90:\rb) {5};
\node[left] at (90+72:\rb) {1};
\node[below left] at (90+2*72:\rb) {2};
\node[below right] at (90+3*72:\rb) {3};
\end{scope}
\end{tikzpicture}
\end{center}
\caption{The dual 4-simplex (in blue) with its dual boundary complex (in red) and boundary triangulation (in black).}\label{penta2}
\end{figure}

For the 4-simplex boundary we have 10 links $[ji]$ in the dual three-complex, to which we associate boundary holonomies $h_{ji}$. There are furthermore 10 faces $[kji]$ (with $k>j>i$) in the dual complex, which we decorate with 2-connection variables $x_{kji}$. For the roots we use the same conventions as in section \ref{sec:5}.

The BFCG amplitude kernel is a product of delta functions imposing 1-flatness  for each  bulk face  and 2-flatness  for each bulk 3-cell in the dual complex. The 10 bulk faces are given by $[ji0]$ and the 10 bulk 3-cells are given by $[kji0]$. The amplitude kernel is thus given by
\begin{align}
Z(\sigma| h_l, x_f) = \int \DD H_{i0} \DD X_{ji0} \; 
\prod_{j>i}\!\! \delta\!\left( H_{0j} h_{ji} H_{i0}\right)  \prod_{k>j>i}\!\!\!\delta\!\left( X_{ji0}+ X_{kj0} -X_{ki0}-H_{0i} \act x_{kji}\right) \, .
\end{align}

We proceed by integrating out the $H_{i0}$ and the $X_{ji0}$ variables.  Firstly we can solve six of the delta functions for the 2-flatness constraints for the six variables $X_{ji0}$ with $i\neq 1$. The remaining four delta functions are given by
\bes\label{2fapp}
&&\delta( H_{01}\act x_{321} -H_{01}\act x_{421} +H_{01}\act x_{431} - H_{02}\act x_{432})\,\times\, \nn\\
&&\delta( H_{01}\act x_{321} -H_{01}\act x_{521} +H_{01}\act x_{531} - H_{02}\act x_{532})\,\times\, \nn \\
&& \delta( H_{01}\act x_{421} -H_{01}\act x_{521} +H_{01}\act x_{541} - H_{02}\act x_{542})\,\times\, \nn\\
&&\delta( H_{01}\act x_{431} -H_{01}\act x_{531} +H_{01}\act x_{541} - H_{03}\act x_{543}) \q.
\ees
These four delta functions are the 2-flatness constraints for the dual 3-cells $[kji1]$ in the boundary. The 2-flatness constraint for $[5432]$ follows from the first four.  We  remain with an integration over the $X_{j10}$ variables, but the path integral does not depend anymore on these variables. This reflects the 2-gauge symmetry and a gauge fixing procedure amounts to dropping the integral over these four variables.

Next we can solve four of the 1-flatness constraints for the four holonomies $H_{20}, \ldots, H_{50}$ and remain with six delta functions
\bes
\delta(h_{12}h_{23}h_{31} ) \, \delta(h_{12}h_{24}h_{41} ) \, \delta(h_{12}h_{25}h_{51} ) \, \delta(h_{13}h_{34}h_{41} ) \, \delta(h_{13}h_{35}h_{51} ) \, \delta(h_{14}h_{45}h_{51} ) \, .
\ees
These are the 1-flatness constraints for the dual faces in the boundary. Note that we only obtain six constraints (for ten dual faces) -- the constraints for the remaining four faces are redundant. We are also left with an integral over $H_{10}$, which reflects the (Lorentz-) gauge invariance of the path integral. 

Using the bulk 1-flatness constraints for the boundary 2-flatness constraints  (\ref{2fapp}) (and dropping in addition a global rotation of the delta-function arguments by $H_{10}$) we can replace the bulk holonomies with boundary holonomies:
\bes
&&\delta(  x_{321} -  x_{421} +  x_{431} - h_{12}\act x_{432})\,
\delta(  x_{321} -  x_{521} +  x_{531} - h_{12}\act x_{532})\,\times\, \nn \\
 &&\delta(  x_{421} -  x_{521} +  x_{541} - h_{12}\act x_{542})\, 
\delta(  x_{431} -  x_{531} +  x_{541} - h_{13}\act x_{543}) \,.
\ees

We are thus left with
\bes\label{ak2}
&&Z(\sigma| h_l, x_f) = \nn\\
&&\;\;\delta(h_{12}h_{23}h_{31} ) \, \delta(h_{12}h_{24}h_{41} ) \, \delta(h_{12}h_{25}h_{51} ) \, \delta(h_{13}h_{34}h_{41} ) \, \delta(h_{13}h_{35}h_{51} ) \, \delta(h_{14}h_{45}h_{51} ) \nn\\
&&\;\;\delta(  x_{321} -  x_{421} +  x_{431} - h_{12}\act x_{432})\,
\delta(  x_{321} -  x_{521} +  x_{531} - h_{12}\act x_{532})\,\times\, \nn \\
 &&\;\;\delta(  x_{421} -  x_{521} +  x_{541} - h_{12}\act x_{542})\, 
\delta(  x_{431} -  x_{531} +  x_{541} - h_{13}\act x_{543})  \q .
\ees
This amplitude kernel features explicitly six of the  ten 1-flatness constraints (for the faces which include the node $1$) and four of the five 2-flatness constraints (for the 3-cells which include the node $1$). These constraints agree with the ones we defined in section \ref{sec:5}.  As mentioned, the remaining 1-flatness constraints and the remaining 2-flatness constraint follow from the ones displayed in (\ref{ak2}). 

We have to include however a cautionary note for the 2-flatness constraint. Not using the 1-flatness constraints, the four 2-flatness constraints in (\ref{ak2}) imply
\bes
h_{12}\act x_{432}-  h_{12}\act x_{532}+ h_{12}\act x_{542}-h_{13}\act x_{543}=0 \q ,
\ees
that is this 2-flatness constraint for the dual 3-face $[5432]$ would be rooted at the node $1$. Multiplying the constraint with $h_{21}$, we still obtain a constraint, which differs by the replacement $h_{21}h_{13} \rightarrow h_{23}$ from the corresponding constraint in section \ref{sec:5}.  That is, here we have also to make use the 1-flatness constraints, to reach a form equivalent to the one used in section \ref{sec:5}.

As outlined in section \ref{sec:BFCG-Yetter}, one can perform a Fourier transformation from the $x$-variables to the $\ell$-variables. Integrating out the $x$-variables, one obtains 6 of the 10 triangle closure constraints. The remaining four follow again from these six constraints. The six constraints are for the triangles which include the vertex $\bar{1}$, but the constraints are based on the node $1$, which is dual to the tetrahedron $[\overline{2345}]$. One needs to use again the 1-flatness constraints to obtain the form of the constraints used in section \ref{sec:5}.

\end{appendix}

\footnotesize
\bibliography{BFCG-Final}
\bibliographystyle{bibstyle_aldo-alpha}

\end{document}